\documentclass[reprint, amsmath, amssymb, aps,superscriptaddress, pra]{revtex4-1}

\usepackage{bm}
\usepackage{dsfont}
\usepackage{natbib}

\usepackage{graphicx}							
\usepackage{dcolumn}							
\usepackage{bm}									
\usepackage{physics}							
\usepackage{amsfonts}							
\usepackage{amssymb}							
\usepackage{tensor}								
\usepackage{dsfont}								
\usepackage{xfrac}								
\usepackage{lipsum}				
\usepackage{xcolor}
\usepackage{tcolorbox}
\usepackage{hyperref}
\usepackage{lineno}   
\usepackage{mathtools}
\usepackage{svg}
\usepackage{appendix}

\usepackage{siunitx}

\makeatletter 
\renewcommand\onecolumngrid{
	\do@columngrid{one}{\@ne}%
	\def\set@footnotewidth{\onecolumngrid}
	\def\footnoterule{\kern-6pt\hrule width 1.5in\kern6pt}%
}

\begin{document}

\title{Error-Resilient Fast Entangling Gates for Scalable Ion-Trap Quantum Processors}

\def\ANU{Department of Quantum Science and Technology, The Australian National University, Canberra, ACT 2601, Australia}

\def\IonQ{IonQ, Inc., College Park, MD, USA}

\author{Isabelle Savill-Brown}
\email{Isabelle.Savill-Brown@anu.edu.au}%
 \affiliation{\ANU}
 \author{Zain Mehdi}
 \affiliation{\ANU}%
    \author{Alexander K. Ratcliffe}
 \affiliation{\IonQ}%
  \author{Varun D. Vaidya}
 \affiliation{\IonQ}%
    \author{Haonan Liu}
 \affiliation{\IonQ}%
  \author{Simon A. Haine}
 \affiliation{\ANU}%
\author{C. Ricardo Viteri}
 \affiliation{\IonQ}%
  \author{Joseph J. Hope}
 \affiliation{\ANU}%
\date{\today}

\begin{abstract}
Non-adiabatic two-qubit gate proposals for trapped-ion systems offer superior performance and flexibility over adiabatic schemes at the cost of increased laser control requirements. Existing fast gate schemes are limited by single-qubit transition errors, which constrain the total number of pulses in high-fidelity solutions. We introduce an improved gate search scheme that enables both local and non-local two-qubit gates in chains containing tens of ions. These protocols use a multi-objective machine design approach that incorporates dominant sources of error in the design to ensure the solutions are compatible with existing fast laser controls. We also generalize previous schemes by allowing for unpaired pulses during the gate evolution. By imposing symmetries on the pulse sequences, we eliminate susceptibility to laser phase noise and further simplify the multi-mode control over the state-dependent motion of the ion crystal. We perform a comprehensive analysis of expected gate performance in the presence of random and systematic experimental errors to demonstrate the feasibility of performing microsecond two-qubit gates between arbitrary ion pairs in current linear ion-trap processors of up to $40$ ions with fidelities approaching $99.9\%$.
\end{abstract}

\maketitle

\section{Introduction}

Trapped ions are one of the most promising platforms for large-scale quantum computation due to their long coherence times \cite{wangSingleIonqubitExceeding2021} and high qubit connectivity \cite{ciracQuantumComputationsCold1995, sorensenEntanglementQuantumComputation2000, garcia-ripollSpeedOptimizedTwoQubit2003}. To date, small-scale trapped ion devices have been used to achieve the highest quantum volume \cite{QuantinuumDominatesQuantum}, the highest fidelities for state-preparation and measurement ($99.99\%$ \cite{anHighFidelityState2022a}), single-qubit gates ($99.99995\%$ \cite{smithSingleQubitGatesErrors2025}), and two-qubit gates ($99.97\%$ \cite{loschnauerScalableHighfidelityAllelectronic2024} for all electronic and $99.94\%$ \cite{clarkHighFidelityBellStatePreparation2021} for laser based). However, scaling to large numbers of ions while maintaining operation speed and fidelity remains the limiting factor to trapped-ion architectures \cite{bruzewiczTrappedIonQuantumComputing2019}.

Strategies for building large-scale trapped-ion quantum computers rely on modular architectures. In each module, ions are entangled via their Coulomb-coupled motion within the trap \cite{leungEntanglingArbitraryPair2018,figgattParallelEntanglingOperations2019,chenBenchmarkingTrappedionQuantum2023,grzesiakEfficientArbitrarySimultaneously2020} and individual traps are connected by physically shuttling ions between modules \cite{mosesRaceTrackTrappedIonQuantum2023, pinoDemonstrationTrappedionQuantum2021, akhtarHighfidelityQuantumMatterlink2023, kaushalShuttlingbasedTrappedionQuantum2020, muraliArchitectingNoisyIntermediateScale2020, mordiniMultizoneTrappedIonQubit2025, sterkClosedloopOptimizationFast2022, schoenbergerShuttlingScalableTrappedIon2025} or long-range photonic interconnects \cite{mainDistributedQuantumComputing2024, monroeLargescaleModularQuantumcomputer2014, knollmannIntegratedPhotonicStructures2024, stephensonHighrateHighfidelityEntanglement2020,drmotaRobustQuantumMemory2023,sahaHighfidelityRemoteEntanglement2024,oreillyFastPhotonMediatedEntanglement2024}. Existing experimental two-qubit entangling gates are performed via spectroscopic excitation of individual modes of collective motion \cite{sorensenEntanglementQuantumComputation2000,gaeblerHighFidelityUniversalGate2016,clarkHighFidelityBellStatePreparation2021, bruzewiczTrappedIonQuantumComputing2019,landsmanTwoqubitEntanglingGates2019}. However, this requires resolution of motional sidebands, which restricts them to being adiabatic relative to the motion of the ions \cite{sorensenEntanglementQuantumComputation2000}. Modern modifications to quasi-adiabatic mechanisms enables gate durations approaching the motional timescale of the crystal~\cite{garcia-ripollCoherentControlTrapped2005,steanePulsedForceSequences2014,palmeroFastPhaseGates2017,sanerBreakingEntanglingGate2023}, although this has only been demonstrated in two-ion systems~\cite{schaferFastQuantumLogic2018,sanerBreakingEntanglingGate2023}. The increased complexity and crowding of the motional spectrum in longer ion chains make error-robust schemes difficult to implement and require longer operation times \cite{landsmanTwoqubitEntanglingGates2019, chenBenchmarkingTrappedionQuantum2023,leungEntanglingArbitraryPair2018, maiScalableEntanglingGates2025,choiOptimalQuantumControl2014} (e.g. Ref.~\cite{chenBenchmarkingTrappedionQuantum2023} achieved a $99.5\%$ gate fidelity in \SI{900}{\micro\second} in a $36$-ion chain). While using fewer ions per module improves gate speed and fidelity, it demands more frequent inter-module links, which create significant resource overhead, namely in clock time and ancilla qubits \cite{bruzewiczTrappedIonQuantumComputing2019, monroeLargescaleModularQuantumcomputer2014, nickersonFreelyScalableQuantum2014, nigmatullinMinimallyComplexIon2016}. Therefore, increasing the number of ions in a single module, while maintaining two-qubit gate speed and fidelity, is key to addressing this scaling problem. 

An alternative approach to entangling ions in longer chains is to use fast gate protocols \cite{bentleyTrappedIonScaling2015, taylorStudyFastGates2017, ratcliffeScalingTrappedIon2018, mehdiFastEntanglingGates2021, mehdiScalableQuantumComputation2020}.
Unlike adiabatic schemes that require long interactions with a narrow-linewidth laser to resolve individual motional sidebands, fast entangling gates use broadband pulses resonant with the qubit splitting to impulsively excite state-dependent motion of the trapped-ion crystal~\cite{garcia-ripollSpeedOptimizedTwoQubit2003, bentleyFastGatesIon2013, duanScalingIonTrap2004, galeOptimizedFastGates2020}. These state-dependent kicks (SDKs) operate outside the Lamb-Dicke regime, meaning gates can theoretically reach GHz two-qubit entangling rates given sufficient laser power and pulse control, fundamentally limited only by the anharmonicity of the trapping potential~\cite{garcia-ripollSpeedOptimizedTwoQubit2003}. Extensive theoretical studies of fast gate schemes have shown that they can achieve high fidelities at MHz rates in scalable 1D and 2D architectures \cite{bentleyTrappedIonScaling2015,taylorStudyFastGates2017, ratcliffeScalingTrappedIon2018, mehdiScalableQuantumComputation2020,mehdiFastEntanglingGates2021}, and are compatible with mixed species crystals \cite{mehdiFastMixedspeciesQuantum2025}. 

However, existing theoretical work is not directly compatible with experimental hardware and largely focuses on the design of gates for two-ion crystals. Extensions to larger ion chains rely on implementing entangling gates much faster than the speed of sound in the trap (the supersonic regime~\cite{savill-brownHighspeedHighconnectivityTwoqubit2025}) \cite{bentleyTrappedIonScaling2015,galeOptimizedFastGates2020, mehdiFastEntanglingGates2021}. Operating on these fast timescales localizes motion to the targeted ions, simplifying motional control, especially in long ion chains. The trade-off is that controlling the delivery of precise laser pulses at these speeds presents a significant challenge for experimental implementation~\cite{bentleyStabilityThresholdsCalculation2016,galeOptimizedFastGates2020,mehdiScalableQuantumComputation2020, mehdiFastEntanglingGates2021}. While SDKs have been successfully demonstrated experimentally \cite{johnsonUltrafastCreationLarge2017, johnsonSensingAtomicMotion2015, putnamImpulsiveSpinmotionEntanglement2024, heinrichUltrafastCoherentExcitation2019}, the more complex control involved in arranging sequences of precisely timed, high-fidelity SDKs has limited current fast gate demonstrations to fidelities of $76\%$ in two-ion systems \cite{wong-camposDemonstrationTwoAtomEntanglement2017}.  
This motivates the need for a comprehensive theoretical analysis and design procedure for fast entangling gates that incorporates experimental limitations and constraints in the control over the multi-mode motion of large trapped-ion crystals using broadband laser pulses. 

In this work we present a machine-design approach for designing fast entangling gate schemes in long linear ion chains that addresses three key challenges in experimental implementations: (1) individual addressing of ions, (2) practical limits to SDK timings (i.e. finite SDK repetition rate), and
(3) compounding motional errors with number of pulses. We demonstrate that high-fidelity gates between local and non-local qubit pairs are achievable in long chains of up to $50$ ions, and further assess the robustness of the presented gate solutions under expected experimental errors. This work is complemented by a companion manuscript~\cite{savill-brownHighspeedHighconnectivityTwoqubit2025}, which establishes the theory of subsonic and supersonic fast gates and demonstrates the viability of `all-to-all' connectivity for MHz-speed quantum information processing in linear ion chains.

The remainder of this article is structured as follows. In \S~\ref{sec:fast_gate_theory}, we describe the fast gate mechanism based on state-dependent momentum kicks (SDKs). In \S~\ref{sec:gate_design}, we outline a method for machine-designing sequences of SDKs that implement high-fidelity entangling gates within technologically motivated constraints using a multi-objective optimization approach. We analyze the performance of these schemes in long ion chains in \S~\ref{sec:results}, with comparison to previous methods for designing SDK sequences. Finally, we discuss the compatibility of these schemes with existing SDK implementations in \S~\ref{sec:SDK_implementation} and investigate their performance in the presence of experimental error sources in \S~\ref{sec:robustness}.

\section{Theory of fast entangling gates based on impulsive SDKs}\label{sec:fast_gate_theory}
\begin{figure}
	\centering
	\includegraphics[width=0.45\textwidth]{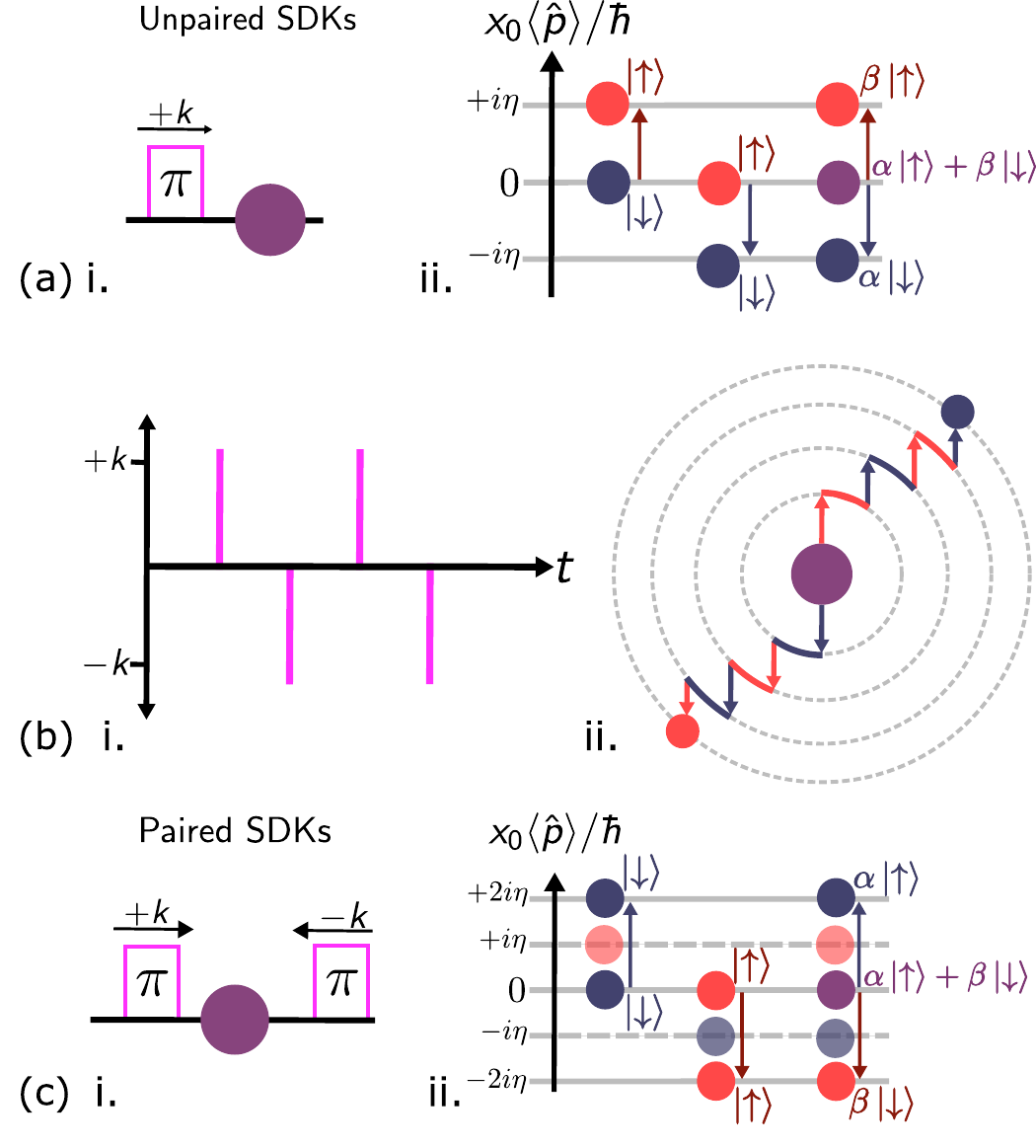}
	\caption{\textbf{State-dependent kicks:} (a) A single SDK is produced by driving an effective $\pi$-rotation of the qubit together with a spin-dependent displacement of $\pm kx_{0} = \pm \eta$ in dimensionless momentum space, where $x_{0} = \sqrt{\hbar/2m\omega_{\rm{COM}}}$.  (b) Sequences of SDKs in the same direction are produced by switching the direction of consecutive pulses (equivalent to changing the sign of $k$). Each SDK is separated by free evolution of the ion motion. (c) Paired SDKs are performed using pairs of counter-propagating $\pi$-pulses. These pulses are separated by a short delay, so we assume no free motional evolution occurs between the forward and counter-propagating pulse. A paired SDK displaces a qubit by $\pm 2kx_{0} = \pm 2\eta$ in dimensionless momentum space and leaves the ion spin unchanged.}
    \label{fig:concept_diagrams1}
\end{figure} 

Fast entangling gates are based on impulsive qubit-state-dependent kicks (SDKs) interspersed with free evolution of the ion motion. These SDKs are implemented using broadband laser pulses that mechanically excite multiple motional modes during the gate (see Fig.~\ref{fig:concept_diagrams2}(b)) rather than spectroscopically addressing individual motional modes \cite{garcia-ripollSpeedOptimizedTwoQubit2003}.

\subsection{State-dependent kicks (SDKs)}
An SDK is performed by driving a $\pi$-rotation in the qubit using a single $\pi$-pulse \cite{heinrichUltrafastCoherentExcitation2019, hussainUltrafastHighRepetition2016, putnamImpulsiveSpinmotionEntanglement2024, guoPicosecondIonqubitManipulation2022} or a sequence of pulses \cite{mizrahiUltrafastSpinMotionEntanglement2013,mizrahiQuantumControlQubits2014,johnsonUltrafastCreationLarge2017, wong-camposDemonstrationTwoAtomEntanglement2017} that are effectively instantaneous compared to the ion motion. The action of a single SDK on the $A$-th ion in an $N$-ion chain is described by the unitary,
\begin{equation}
\label{eq:SDKUnitary_PositionBasis}
    \hat{U}_{\rm{SDK}}^{(A)} = \hat{\sigma}_{x}^{(A)}e^{i(k \hat{x}^{(A)} + \phi^{(A)})\hat{\sigma}_{z}^{(A)}}~,
\end{equation}
where $k$ and $\phi$ are the wavenumber and phase of the laser. The result is an inversion of the qubit spin-state, and a spin-dependent momentum transfer of $\pm \hbar k$ as shown in Figure~\ref{fig:concept_diagrams1}(a). 

Expanding the position operator into the normal mode basis, $k\hat{x}^{(A)} = \sum_{\alpha = 1}^{N} b_{\alpha}^{(A)} \eta_{\alpha}(\hat{a}_{\alpha}+\hat{a}_{\alpha}^{\dagger})$ with $\eta_{\alpha} = k\sqrt{\hbar/(2m\omega_{\alpha})}$ as the mode-dependent Lamb-Dicke parameter, we can write the action of an SDK as
\begin{align}
\label{eq:SDKUnitary_ModeBasis}
    \hat{U}_{\rm SDK}^{(A)} = \hat{\sigma}_x^{(A)}e^{i\phi^{(A)}\hat{\sigma}_z^{(A)}}\prod_{\alpha = 1}^N \hat{D}_\alpha(ib_\alpha^{(A)}\eta_\alpha \hat{\sigma}_z^{(A)}) \,,
\end{align}
where $\hat{D}_{\alpha}(\beta) = \exp\left[\beta\hat{a}_{\alpha}^{\dagger}-\beta^{*}\hat{a}_{\alpha}\right]$ is the displacement operator for the $\alpha$-th motional mode. Here we allow $\beta = ib_{\alpha}^{(A)}\eta_{\alpha} \hat{\sigma}_{z}^{(A)}$ to be an operator acting on the qubit state. In the above expression, we have assumed excitation of the ion chain motion along a single principal axis of the trap such that $N$ modes are excited by a single SDK.

\subsection{Implementing a $\sigma_z\otimes\sigma_z$ entangling operation}

Here we describe how a sequence of SDKs addressing ions $A$ and $B$ interspersed by periods of free motional evolution can realise a $\sigma_z^{(A)}\otimes\sigma_z^{(B)}$-type entangling operation \cite{garcia-ripollSpeedOptimizedTwoQubit2003}. It is convenient to first move to the interaction picture with respect to the motional Hamiltonian:
\begin{align}
    \hat{U} \rightarrow e^{i\sum_\alpha \hat{a}^\dag_\alpha \hat{a}_\alpha \omega_\alpha t} \hat{U} e^{-i\sum_\alpha \hat{a}^\dag_\alpha \hat{a}_\alpha \omega_\alpha t} \,,
\end{align}
which is equivalent to moving into the rotating frame of each motional mode, i.e. $\hat{a}_\alpha(t) = \hat{a}_\alpha(0)e^{-i\omega_\alpha t}$. If we then take the SDK to be effectively instantaneous with respect to the motion of the ion chain, the unitary operator for a fast gate operation can be written as the time-ordered product of $\mathcal{N}$ displacement operations at times $\{t_1,\dots, t_\mathcal{N}\}$ using Eq.~\eqref{eq:SDKUnitary_ModeBasis}:
\begin{align}
\label{eq:GateUnitaryFull}
    \hat{U}_{\rm G} =\mathcal{T}\left[\prod_{m = A,B} \prod_{\alpha = 1}^{N}\prod_{j=1}^{\mathcal{N}}\right.&\left.\hat{\sigma}_{x}^{(m)}e^{i\kappa_{j}\phi^{(m)}(t_{j})\hat{\sigma}_{z}^{(m)}} \right.\notag\\ \times& \left.\hat{D}_{\alpha}\left(i\kappa_{j}\eta_{\alpha}b_{\alpha}^{(m)}\hat{\sigma}_{z}^{(m)}e^{i\omega_{\alpha} t_{j}} \right)\right] \,.
\end{align}
Here we have allowed the direction of each SDK to switch between $k$ and $-k$ through the introduction of the variable $\kappa_j = \pm 1$, noting that reversing the SDK direction also changes the sign of the laser phase (i.e. $k\rightarrow -k$, $\phi\rightarrow -\phi$). This unitary describes (1) $\mathcal{N}$ single-qubit rotations on each targeted ion, with rotation angle $\kappa_{j}\phi^{(m)}(t_j)$ around the $\hat{\sigma}_z$ axis of the Bloch sphere; (2) $\mathcal{N}$ spin flips ($\hat{\sigma}_x$ operations) on each qubit; and (3) qubit-state-dependent displacements of each motional mode. 

The mechanism behind two-qubit entanglement in the unitary given by Eq.~\eqref{eq:GateUnitaryFull} is the non-commutativity between displacements along different axes in phase-space, i.e. $\hat{D}(\alpha)\hat{D}(\beta) = e^{i\theta}\hat{D}(\alpha+\beta)$ where $\theta = \rm{Im}[\alpha\beta^*]$. As the displacement amplitudes in Eq.~\eqref{eq:GateUnitaryFull} are qubit-state-dependent, this leads to a two-qubit (2Q) state-dependent phase which can be explicitly shown by rewriting Eq.~\eqref{eq:GateUnitaryFull} following the derivation in Appendix~\ref{sec:unitary_derivation_appendix}~,
\begin{align}
\label{eq:GateUnitary_PhaseGate}
\hat{U}_{\rm G} =  &\left(\hat{\sigma}_x^{(A)}\hat{\sigma}_x^{(B)}\right)^\mathcal{N}e^{i\Theta_{\rm 2Q}\hat{\sigma}_{z}^{(A)}\hat{\sigma}_{z}^{(B)}} e^{i(\Phi_{\rm 1Q}^{(A)}\hat{\sigma}_z^{(A)}+\Phi_{\rm 1Q}^{(B)}\hat{\sigma}_z^{(B)})}  \notag \\   &\qquad\times \prod_{\alpha}\hat{D}_\alpha\left(\Delta \beta_\alpha[b_\alpha^{(A)}\hat{\sigma}_z^{(A)}+b_\alpha^{(B)}\hat{\sigma}_z^{(B)}]\right) \,,
\end{align}
where the 2Q phase is given by
\begin{align}
    \Theta_{\rm 2Q} =2\sum_{\alpha = 1}^{N}\eta_{\alpha}^{2} b_{\alpha}^{(A)}b_{\alpha}^{(B)}\sum_{k\neq j} ^{\mathcal{N}}\kappa_{j}\kappa_{k}(-1)^{j+k}\sin\left(\omega_{\alpha}(t_{j} - t_{k})\right) \,,\label{eq:entangling_phase} 
\end{align}
and the residual state-dependent displacement for the $\alpha$-th normal mode is given by
\begin{align}
    \Delta \beta_{\alpha} &= i\eta_{\alpha}\sum_{j=1}^{\mathcal{N}} \kappa_{j}(-1)^{j+1} e^{i\omega_{\alpha} t_{j}}~.\label{eq:motional_displacement}
\end{align} 
Note that the factor of $(-1)^j$ arises from the spin flip associated with each SDK, as shown in Appendix~\ref{sec:unitary_derivation_appendix}. We also note that the final 2Q phase has no dependence on the laser phase $\phi(t)$. Instead, this only contributes a single-qubit phase, 
\begin{align}
    \Phi_{\rm{1Q}} =\sum_{j = 1}^{\mathcal{N}}(-1)^{j+1}\kappa_{j}\phi(t_{j}) ~,\label{eq:laser_phase}
\end{align}
which we discuss in \S~\ref{sec:robustness}.

Provided the total number of SDKs ($\mathcal{N}$) is even such that we can take $(\sigma_x)^\mathcal{N}=1$, this will lead to a $\sigma_z\otimes \sigma_z$-type entangling phase of the form $\hat{U}_{ZZ}(\Theta_{\rm{2Q}}) = \exp(i\Theta_{\rm{2Q}}\hat{\sigma}_z^{A}\otimes\hat{\sigma}_z^{B})$. In the following section we will consider a gate optimization protocol that ensures this condition is satisfied for all gate solutions considered in this work.

\subsubsection*{Fast gate fidelity}
\begin{figure*}
	\centering
	\includegraphics[width=0.85\textwidth]{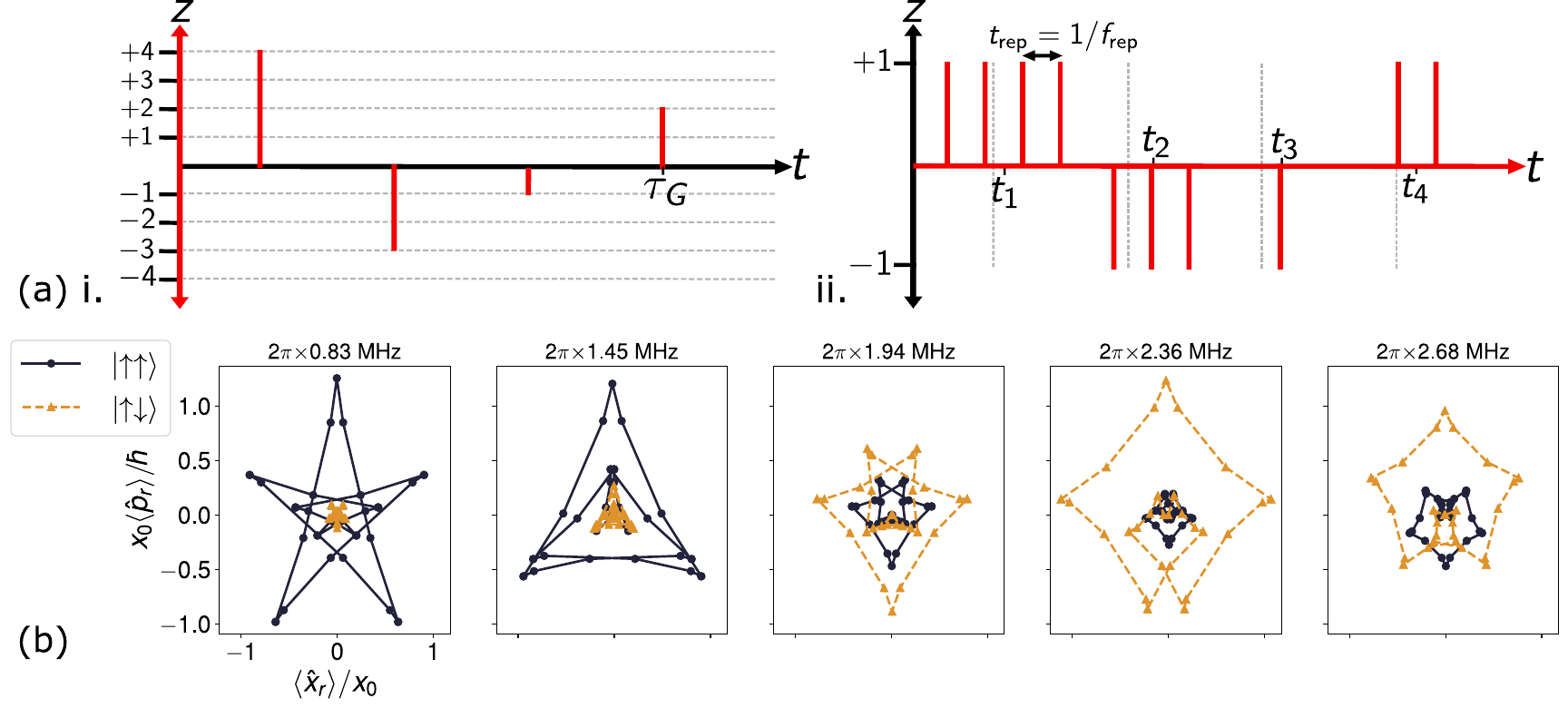}
	\caption{\textbf{Gate scheme design:} (a) Schematic diagram of the multi-stage optimization process we use to design gate schemes that ensure the correct two-qubit phase is accumulated and the spin-motional degrees of freedom are disentangled at the end of the gate. In the first stage (a.i), we assume all SDKs arrive simultaneously and optimize the magnitude and direction of each group. In the second stage (a.ii), groups are separated into their constituent SDKs based on a finite SDK repetition rate, $t_{\rm{rep}}$, and the timing of each group is optimized. (b) The simulated rotating phase-space trajectories of the axial motional modes in a $5$-ion chain during a gate between a pair of ions at the edge of the chain for an optimized solution with time $0.89 \rm{\mu s}$ ($0.74$ center-of-mass periods) and theoretical gate error $\varepsilon_{\rm{av}} = 2.5\times 10^{-4}$. Each mode is rotating at its normal frequency, with the lowest frequency corresponding to the center-of-mass mode. We have only shown the trajectories for the $\ket{\uparrow \uparrow}$ and $\ket{\uparrow \downarrow}$ two-qubit spin-states as the trajectories for the $\ket{\downarrow \downarrow}$ and $\ket{\downarrow \uparrow}$ states are given by symmetry, $(\hat{x}_{r}, \hat{p}_{r}) \rightarrow (-\hat{x}_{r}, -\hat{p}_{r})$. The total entangling phase accumulated during the gate originates from the difference in areas enclosed by the $\{\ket{\uparrow \uparrow},\ket{\downarrow \downarrow}\}$ and $\{\ket{\uparrow \downarrow}, \ket{\downarrow \uparrow}\}$ trajectories across all the modes.}
    \label{fig:concept_diagrams2}
\end{figure*} 

Assuming the spin-motional degrees of freedom are disentangled by the end of the gate operation (i.e. $\Delta \beta_\alpha= 0$), Eq.~\eqref{eq:GateUnitary_PhaseGate} realises a maximally-entangling phase gate when $\Theta_{\rm 2Q}=\pi/4$, up to single-qubit rotations. Geometrically, this is equivalent to creating closed trajectories in phase space, such that the difference in the area enclosed by the $\{\ket{\uparrow\uparrow},\ket{\downarrow\downarrow}\}$ and $\{\ket{\uparrow\downarrow},\ket{\downarrow\uparrow}\}$ states is $\frac{\pi}{2}$-- see Fig.~\ref{fig:concept_diagrams2}(b).

The degree to which these conditions can be satisfied will determine the theoretical fidelity of the fast entangling gate operation. We use the qubit-state-averaged fidelity of the unitary Eq.~\eqref{eq:GateUnitary_PhaseGate} as compared to the ideal unitary $\hat{U}_{\text{id}} = \hat{U}_{\rm ZZ}(\pi/4)$, 
\begin{align}
	F_{\rm{av}} = \frac{1}{\int_{\ket{\psi_{0}}}d\ket{\psi_{0}}}\int_{\ket{\psi_{0}}}d\ket{\psi_{0}}&\text{Tr}_{m}\left[\bra{\psi_{0}}\hat{U}^{\dagger}_{\text{id}} \hat{U}_{\text{G}}\ket{\psi_{0}}\right.\notag\\&\left.\bra{\psi_{0}}\otimes \rho_{m} \hat{U}_{\text{G}}^{\dagger}\hat{U}_{\text{id}}\ket{\psi_{0}}\right]~,
	\label{eq:full_state_averaged_fidelity}
\end{align}
where the partial trace is taken over the motional states, assuming an initial thermal product state.

Assuming errors due to imperfect 2Q phase accumulation ($\Delta \Theta = \Theta_{\rm{2Q}} - \frac{\pi}{4}$) and residual spin-motional entanglement ($\Delta \beta_{\alpha}$) are small, we can Taylor expand Eq.~\eqref{eq:full_state_averaged_fidelity} to obtain a simple expression for the theoretical gate error $\varepsilon_{\rm{av}} = 1- F_{\rm{av}}$ \cite{galeOptimizedFastGates2020, mehdiFastEntanglingGates2021},
\begin{align}
    \varepsilon_{\rm{av}} = \frac{2}{3}& \left\lvert\Delta \Theta\right\rvert^{2}\notag\\& + \frac{4}{3} \sum_{\alpha=1}^{N} \left(\frac{1}{2} + \bar{n}_{\alpha}\right)\left[(b_{\alpha}^{(A)})^{2} + (b_{\alpha}^{(B)})^{2}\right] \left\lvert\Delta \beta_{\alpha}\right\rvert^{2}~,\label{eq:infidelity}
\end{align}
where $\bar{n}_{\alpha} = \left( e^{\frac{\hbar \omega_{\alpha}}{k_{B}T}} -1\right)^{-1}$ is the average phonon occupation of each mode in terms of the temperature of the system, $T$. Note that for perfect motional restoration ($\Delta \beta_{\alpha} = 0$), the fast gate mechanism is insensitive to the system temperature. In our results we assume a constant temperature of \SI{30}{\micro\kelvin}, however, we discuss the impact of higher system temperatures when there is imperfect motional restoration in \S~\ref{sec:robustness}. This temperature corresponds to a center-of-mass mode occupation of $\bar{n}_{\rm{COM}} = 0.36$ ($4.4$) in a chain of $5$ ($50$) ions with a minimum interion separation of \SI{3}{\micro\meter}.

\section{Multi-objective fast gate optimization}\label{sec:gate_design} 

In this manuscript, we focus on three key metrics of gate performance: the theoretical gate fidelity ($1-\mathcal{\varepsilon_{\rm{av}}}$), the gate time ($\tau_{G}$), and the number of SDKs involved in the gate ($\mathcal{N}$). In addition to these metrics, we also quantify the effective SDK repetition rate, $f_{\rm{rep}}$, (the minimum timing separation between consecutive SDKs) required in the schemes we consider. Both the number of SDKs and the repetition rate are indicative of the laser control requirements and are therefore a critical consideration in designing fast gate schemes for near-term experimental implementations. 

Given the complexity of the search space of sequences of SDKs that implement a high-fidelity fast gate, we take a heuristic approach where we machine design fast gate schemes using a multi-stage optimization of a generalised anti-symmetric gate (APG) scheme adapted from Ref.\cite{galeOptimizedFastGates2020, mehdiFastEntanglingGates2021}. This method is better-suited fast gate design in more complex systems where more motional control is required such as long ion chains compared to other schemes \cite{garcia-ripollSpeedOptimizedTwoQubit2003,bentleyFastGatesIon2013, duanScalingIonTrap2004} which impose stronger constraints based on the dynamics in two-ion systems. An anti-symmetric gate scheme guarantees an even number of SDKs, which, as discussed in the previous section, is necessary to ensure the qubits are returned to their original spin states at the end of the gate.  Furthermore, the anti-symmetry of these schemes also guarantees the momentum restoration of each motional mode, which is advantageous in larger-ion systems where there are a larger number of motional modes that must be disentangled from the two-qubit spin state at the end of the operation \cite{galeOptimizedFastGates2020}.

This scheme considers $K$ groups SDKs where the $j$-th group is composed of $z_{j}$ SDKs arriving at timings centered around $t_{j}$,
\begin{align}
	\bm{z} &= \{-z_{K/2},...-z_{2},-z_{1}, z_{1}, z_{2}, ..., z_{K/2}\}~,\\
	\bm{t} &= \{-t_{K/2}, ..., -t_{2}, -t_{1}, t_{1},t_{2},..., t_{K/2}\} ~.
\end{align}
The sign of $z_{j}$ corresponds to the direction of the SDKs in the $j$-th pulse group, which depends on the direction of the beam $\kappa_{j}$, and the number of spin flips (and thus SDKs) that have occurred during the gate: $(-1)^{\sum_{k=-K/2-1}^{j-1}|z_{k}|+1}\kappa_{j}$. 

We use a two-stage optimization over $\bm{z}$ and $\bm{t}$ using the theoretical gate error (Eq.~\eqref{eq:infidelity}) as our cost function such that the phase mismatch, 
\begin{equation}\label{eq:entangling_phase_condition}
    \Delta\Theta = \left\lvert2\sum_{\alpha = 1}^{N}\eta_{\alpha}^{2} b_{\alpha}^{(A)}b_{\alpha}^{(B)}\sum_{k\neq j} ^{\mathcal{N}}z_{j}z_{k}\sin\left(\omega_{\alpha}(t_{j} - t_{k})\right)\right\rvert - \frac{\pi}{4}
\end{equation}
and the residual spin-motional entanglement, 
\begin{equation}
     \Delta \beta_{\alpha} = i\eta_{\alpha}\sum_{j=1}^{\mathcal{N}} z_{j}\sin(\omega_{\alpha} t_{j})~,\label{eq:motional_displacement_condition_APG}
\end{equation}
are minimized. We note that this is equivalent to Eq.~\eqref{eq:motional_displacement} under the anti-symmetric constraint of the APG scheme. Each optimization is performed over an undersampled, high-dimensional, and non-convex search space. As a result, we do not expect to find the global minimum in our searches. However, we find that the observed trends in our results are robust to increases in the search density. This procedure is outlined in Figure~\ref{fig:concept_diagrams2}(a).
\begin{figure*}
	\centering
	\includegraphics[width=\textwidth]{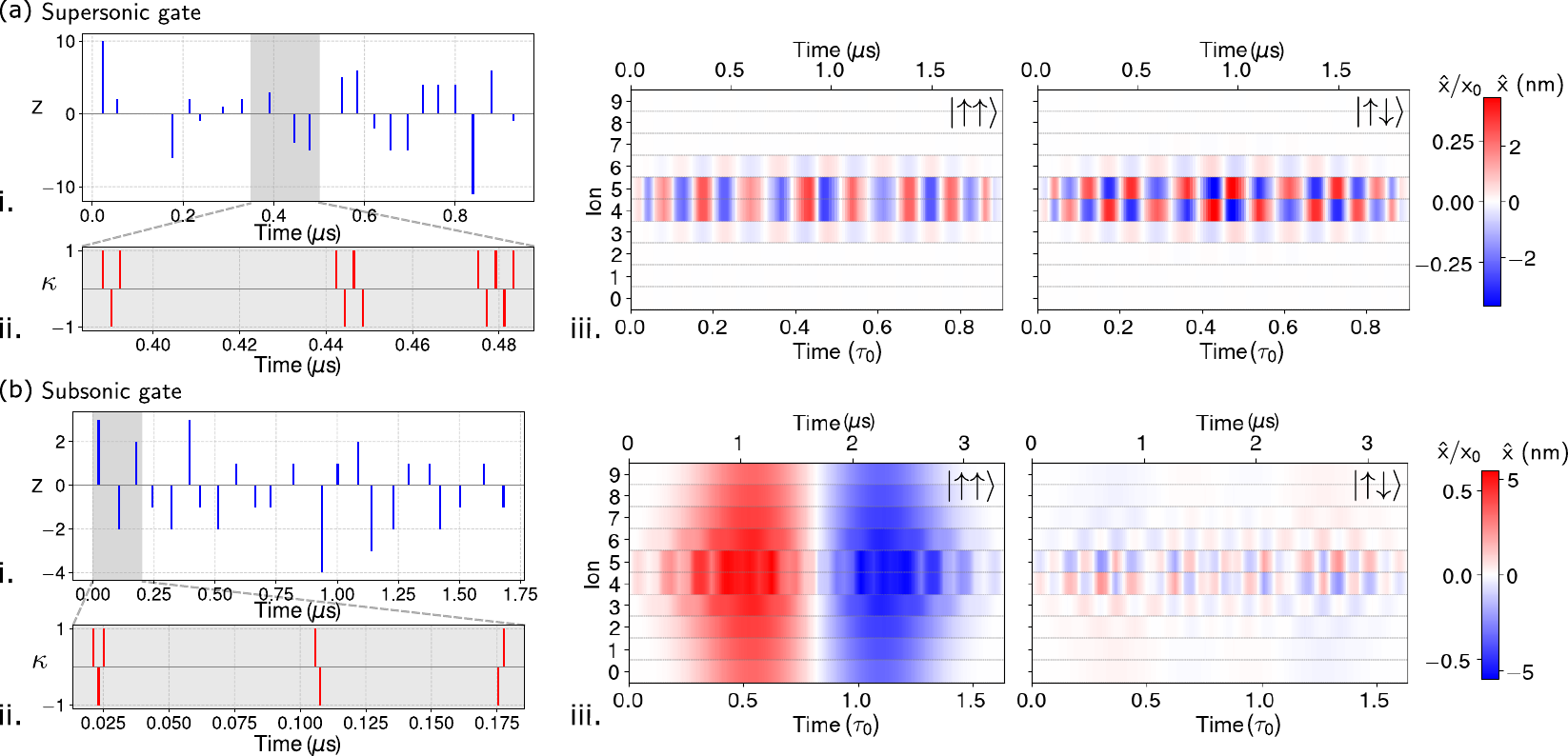}
	\caption{\textbf{Example optimized fast gate schemes:} Optimized gate solutions for (a) a supersonic gate and (b) a subsonic gate between ions $4$ and $5$ in a $10$-ion chain, assuming a \SI{500}{\mega\hertz} SDK repetition rate. Both gate schemes are anti-symmetric around $t=0$. (a.i) A supersonic gate scheme composed of $178$ SDKs. The magnitude and sign of $z$ indicates the number and direction of the SDKs in each group. The timing of the SDKs in each group is centered around the time shown, with consecutive SDKs separated according to the SDK repetition rate. The pulse sequence (with this repetition rate included) that realizes three of these groups is shown in (a.ii). Within each group, consecutive pulses switch direction, with the direction of the first pulse in each group chosen based on the number of spin-flips (SDKs) that have already occurred. (a.iii) The simulated motional trajectories of the ions during this sequence of SDKs.  We have only shown the trajectories for the $\ket{\uparrow \uparrow}$ and $\ket{\uparrow \downarrow}$ two-qubit spin-states as the trajectories for the $\ket{\downarrow \downarrow}$ and $\ket{\downarrow \uparrow}$ states are given by symmetry, $x \rightarrow -x$. The same is shown in (b) for a subsonic gate made up of $78$ SDKs.}
    \label{fig:trajectories}
\end{figure*} 
\subsection{Optimization over number of SDKs}
In the first stage,
a global optimization is performed over $\bm{z}$ while the timings are fixed based on a desired gate time $\tau_{G}$: $\bm{t} = \frac{\tau_{G}}{K}\{-\frac{K}{2}, ..., -2, -1, 1,2,..., \frac{K}{2}\}$. A global optimization over $\bm{z}$ is advantageous as the cost function is quartic in the solution space ($\varepsilon_{\rm{av}} \sim O(z_{k}^{4})$), which is significantly computationally cheaper than optimizing $\bm{t}$, which varies sinuisoidally in the solution space. To further increase computational efficiency in this stage, we assume there is no free evolution between SDKs in a single group.

We perform local gradient-descent optimizations from a large ensemble ($10^{3}$) of randomly sampled initial conditions within the allowed parameter space, consisting of $30-50$ SDK groups with $|z_{j}| \leq 5$ SDKs per group. After each local minimization the continuous solution is projected onto the nearest integer SDK configuration, and the best candidate is passed to the subsequent optimization stage. We note that we use a larger search dimensionality with fewer SDKs per group compared to previous work~\cite{galeOptimizedFastGates2020, mehdiFastEntanglingGates2021} (which considered $16-20$ groups with $|z_{j}| \leq10$). While increasing the dimensionality reduces the search density of the solution space, we find the additional timing degrees of freedom are better suited for identifying high-fidelity solutions with fewer SDKs.

Previous work has identified that 2Q gate errors compound with the number of imperfect SDKs -- specifically, the achievable fidelity is bounded by $F \geq (1-2\mathcal{N} \epsilon_{\pi})F_{0}$, for a gate with ideal fidelity $F_{0} = 1-\varepsilon_{\rm{av}}$ and $\epsilon_{\pi}$ is the characteristic population transfer error between spin states \cite{bentleyStabilityThresholdsCalculation2016, mehdiFastEntanglingGates2021}. As SDK errors have been identified as the primary technical limitation to implementing 2Q gates \cite{wong-camposDemonstrationTwoAtomEntanglement2017, mehdiFastEntanglingGates2021, galeOptimizedFastGates2020}, we develop schemes that minimize the gate error with the fewest number of SDKs by using a modified cost function in this stage that includes SDK errors,
\begin{equation}
    \varepsilon_{\epsilon} = 2\epsilon_{\pi}\sum_{k=1}^{K/2}|z_{k}|\varepsilon_{\rm{av}}~,\label{eq:SDK_errors}
\end{equation}
where we choose SDK errors in the range of current experimental demonstrations ($\epsilon_{\pi} = 10^{-2}$ \cite{johnsonSensingAtomicMotion2015}) to the spontaneous emission limit of the candidate ion used in our optimizations, $^{133}$Ba ($\epsilon_{\pi} = 10^{-7}$  \cite{liuHighFidelityRamanSpinDependent2025}). We further discuss the impacts of imperfect SDKs on gate performance in \S~\ref{sec:robustness}.

\subsection{Optimization over SDK timings}
The second optimization stage then uses the optimal solutions for $\bm{z}$ from the first stage and performs local optimizations on $\bm{t}$ to further refine the gate solution. In this second stage, the constituents of each pulse group are broken up into individual SDKs and enforce a minimum timing separation on the SDKs, $t_{j+1} = t_{j} + 1/f_{\rm{rep}}$, for a specific SDK repetition rate, $f_{\rm{rep}}$. Unlike previous work \cite{ galeOptimizedFastGates2020, mehdiFastEntanglingGates2021}, we do not enforce the SDK timings to be integer multiples of this repetition rate. 

\section{Optimization results}\label{sec:results}
\begin{figure*}
	\centering
	\includegraphics[width=\textwidth]{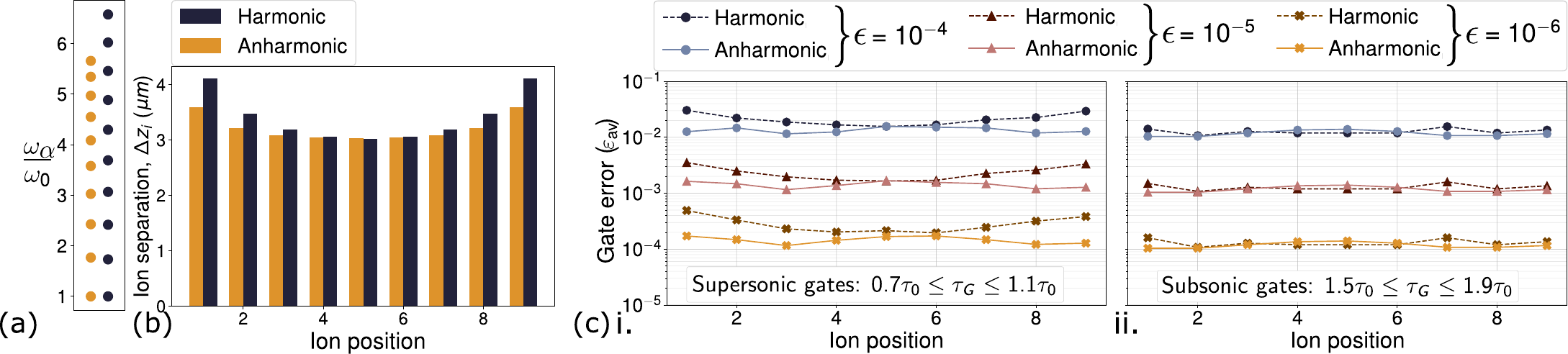}
	\caption{\textbf{10-ion chain in an anharmonic and harmonic trapping potential:} $10$-ion chain with a minimum ion separation of \SI{3}{\micro\meter} in a harmonic trapping potential compared to an anharmonic trapping potential. The harmonic trap has a trapping frequency $\omega_{t} =$ \SI{2.62}{\mega\hertz}, and the anharmonic trap is characterized by $\kappa_{2} = 9.38\times 10^{-13}$ \unit{\joule\per\meter\squared} and $\kappa_{4} = 5.15\times 10^{-3}$ \unit{\joule\per\meter\tothe{4}}. (a) The normal motional mode frequency structure for each trap relative to the center-of-mass frequency $\omega_{\rm{COM}} \equiv  \omega_{0}$. (b) The separation between adjacent ions as a function of the position in the chain. (c) Optimized gate solutions for gates between neighboring ions at different positions in the chain. Gate error for various SDK errors is plotted as a function of the position of the ion pair in the chain when using a harmonic trap or a quartic trap. Each gate was optimized for a \SI{500}{\mega\hertz} SDK repetition rate and gate times after optimization were in the range (c.i) $0.7 - 1.1$ center-of-mass oscillation periods and (c.ii) $1.5-1.9$ center-of-mass oscillation periods.}
	\label{fig:quartic_harmonic_results}
\end{figure*}

In this section, we investigate the performance of our machine-designed fast gate schemes in ion chains with up to $50$-ions. We report gate times in terms of the center-of-mass (COM) oscillation period, $\tau_{0} = 2\pi/\omega_{\rm{COM}}$, as this characterizes the timescale of motional dynamics in the system. 

In our analysis, we consider gate performance in two time regimes: the supersonic regime and the subsonic regime. These regimes are defined based on the timescale for phonon-mediated entanglement in large-ion crystals, $\tau_{\rm{travel}} \approx \tau_{0}/(\omega_{BR}/\omega_{COM} - 1)$ where $\omega_{\rm COM}$ and $\omega_{\rm BR}$ are the center-of-mass and breathing (stretch) mode frequencies, respectively \cite{savill-brownHighspeedHighconnectivityTwoqubit2025}. We show an example of the SDK schemes and motional dynamics of a $10$-ion chain for optimized solutions in each of these regimes in Figure~\ref{fig:trajectories}. Supersonic fast gates ($\tau_{G} < \tau_{\rm{travel}}$) are characterized by highly localized ion dynamics, which can be seen in Fig.~\ref{fig:trajectories}(a.iii). As a result, supersonic fast gates are limited to nearest-neighbor ion pairs \cite{bentleyTrappedIonScaling2015}. In contrast, subsonic gates ($\tau_{G} > \tau_{\rm{travel}}$) exhibit collective motion of the entire ion chain (see Fig.~\ref{fig:trajectories}(b.iii)), which supports fast gates between both local and non-local ion pairs \cite{savill-brownHighspeedHighconnectivityTwoqubit2025}.

The fast gate mechanism is agnostic to the ion species and qubit type beyond the ability to perform high-fidelity state-dependent momentum kicks. In our model, we consider a linear chain of $^{133}$Ba ions in an axial trapping potential of the form: $V_{\rm{trap}}(z) = \kappa_{2}z^{2}/2 + \kappa_{4}z^{4}/4$ \cite{liHomogeneousLinearIon2022, linLargescaleQuantumComputation2009}. The trapping parameters ($\kappa_{2}$, $\kappa_{4}$) are chosen to provide a minimum ion separation of \SI{3}{\micro\meter} (see Appendix~\ref{sec:trap_appendix}) which is sufficient for individual ion addressing \cite{manovitzTrappedIonQuantumComputer2022, chenBenchmarkingTrappedionQuantum2023, johnsonUltrafastCreationLarge2017, pogorelovCompactIonTrapQuantum2021}. We consider SDKs driven by counter-propagating \SI{532}{\nano\meter} Raman beams coupled to the axial modes of the targeted ions. We assume these beams are equally tilted from the RF null of the trap at an angle of $\theta = \frac{\pi}{3}$ to enable individual addressing of the target ions such that the magnitude of each kick is $\hbar k$ in the axial direction (see Fig.~\ref{fig:RamanBeamGeometry}(c)). The experimental feasibility of SDK implementations in this system are discussed in \S~\ref{sec:SDK_implementation}.

\subsection{Comparison to previous gate designs}
To demonstrate the improvements made to our gate design, we compare our optimized gate solutions with previous approaches, where ion chains were modeled in a purely harmonic potential and SDKs were implemented using counter-propagating $\pi$-pulse pairs.

\subsubsection{Anharmonic trapping potential} 
Anharmonic potentials have been used in trapped-ion experiments to confine large linear chains of ions with equidistant separations \cite{paganoCryogenicTrappedIonSystem2018, cetinaControlTransverseMotion2022, leungEntanglingArbitraryPair2018}, however, previous theoretical fast gate work has only considered harmonic axial potentials \cite{bentleyTrappedIonScaling2015, mehdiFastEntanglingGates2021}. 

In Figure~\ref{fig:quartic_harmonic_results}, we compare the performance of gates optimized between local ion pairs in a $10$-ion chain in an anharmonic potential and a harmonic potential. The trapping parameters ($\omega_{t}$ for the harmonic potential and $(\kappa_{2}, \kappa_{4}$) for the anharmonic potential) were chosen to provide a \SI{3}{\micro\meter} minimum ion separation (see Fig.~\ref{fig:quartic_harmonic_results}(b)). To present an equivalent comparison of gates with different numbers of SDKs, we present the total gate error with SDK errors ($\epsilon$) included using Eq.~\eqref{eq:SDK_errors}.  

In the supersonic regime (Fig.~\ref{fig:quartic_harmonic_results}(c.i)), gate solutions between ions near the edge of the chain perform worse in the harmonic trap, with up to a factor of five increase in the gate error compared to gates optimized in the anharmonic trapping potential. Supersonic gates are mediated by the local couplings between ions, which decay with the cube of the ion separation \cite{bentleyTrappedIonScaling2015}. In the center of the chain, the ion separation is the same in both potentials and therefore there is no significant difference in the gate performance. However, ion separation increases towards the edge of the chain, as shown in Fig.~\ref{fig:quartic_harmonic_results}(b), leading to weaker couplings between ions. This is more significant in the harmonic trap, where the ion separation increases to \SI{4.1}{\micro\meter} at the end of the chain, compared to \SI{3.59}{\micro\meter} in the anharmonic potential. Therefore, at the edge of the chain, more SDKs are needed to compensate for the weaker coupling between ions in the harmonic trap, so SDK errors have a greater impact on the total gate error. 

Gates optimized in the subsonic regime (Fig.~\ref{fig:quartic_harmonic_results}(c.ii)) do not exhibit the same variations in performance as observed in the supersonic regime. The gate dynamics in this regime involve the collective motion of the entire ion chain and therefore are independent of the coupling between individual ion pairs.

\begin{figure}
	\centering
	\includegraphics[width=0.5\textwidth]{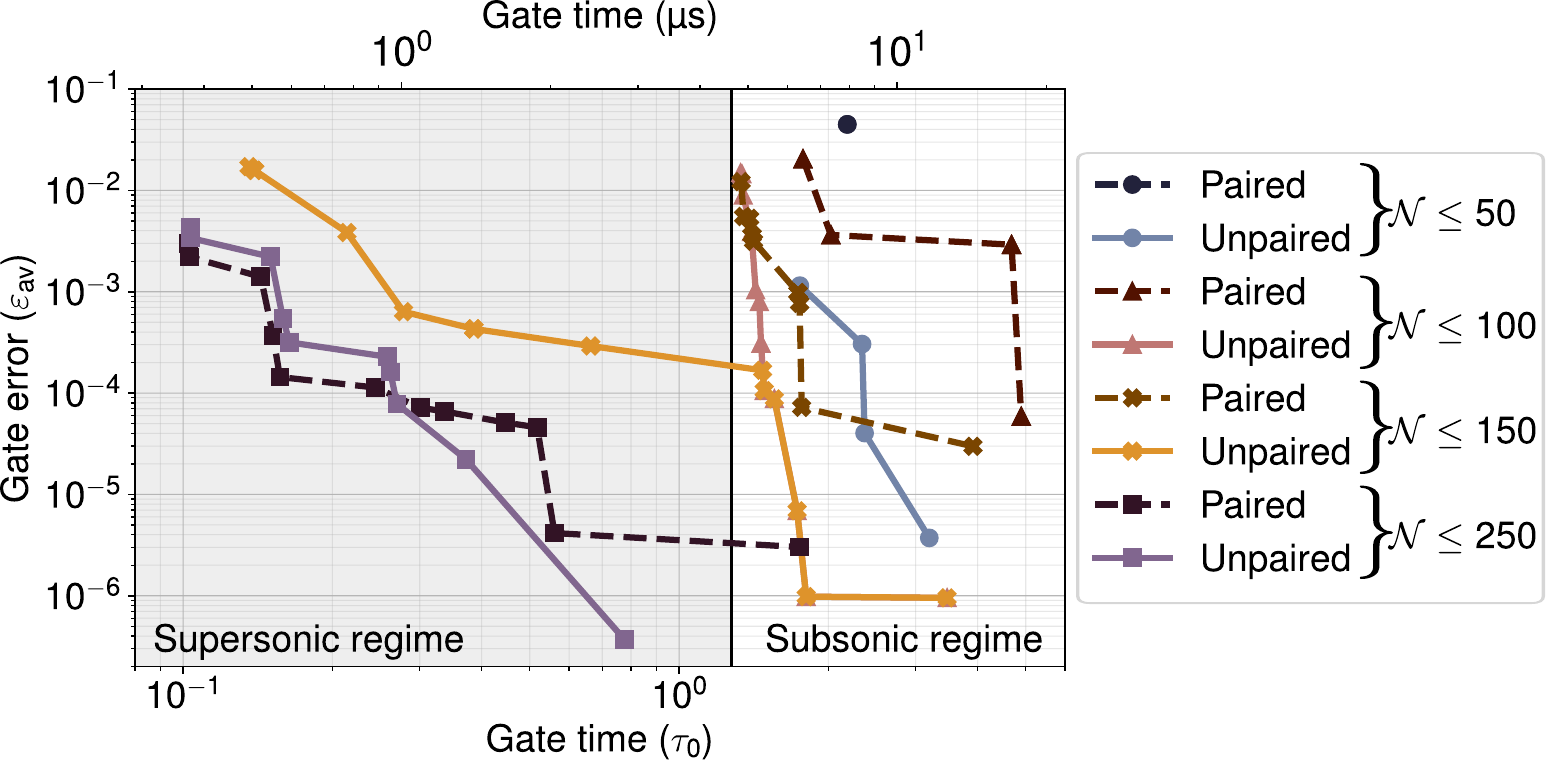}
	\caption{\textbf{Gate performance with paired and unpaired SDKs:}  Gate error (in the absence of pulse errors) is shown for varying gate times with an upper-bound placed on the number of SDKs ($\mathcal{N}$) allowed per gate in schemes. Gate performance is compared for schemes optimized with SDKs implemented using individual (unpaired) $\pi$-pulses compared to counter-propagating $\pi$-pulse pairs. To fairly compare the schemes due to the difference in laser requirements needed to produce an SDK in the paired and unpaired schemes, we apply the following conditions: $2f_{\rm{rep, paired}} = f_{\rm{rep, unpaired}}$ and $\mathcal{N}_{\rm{paired}} =2 \mathcal{N}_{\rm{unpaired}}$. The unpaired scheme always outperforms the paired scheme in the subsonic (unshaded) regime. All gate solutions assume a \SI{500}{\mega\hertz} unpaired (\SI{250}{\mega\hertz} paired) SDK repetition rate for $2$ neighboring ions at the edge of a $20$-ion chain in an anharmonic trap characterized by $\kappa_{2} = 2.78\times 10^{-13}$ \unit{\joule\per\meter\squared} and $\kappa_{4} = 4.27\times 10^{-4}$ \unit{\joule\per\meter\tothe{4}}.}
	\label{fig:unpaired_plot}
\end{figure} 

\subsubsection{SDKs with individual $\pi$-pulses}
Previous theoretical work on fast gate design has exclusively considered fast entangling gates based on paired SDKs implemented using counter-propagating $\pi$-pulses \cite{garcia-ripollSpeedOptimizedTwoQubit2003,bentleyFastGatesIon2013,bentleyTrappedIonScaling2015, taylorStudyFastGates2017,ratcliffeScalingTrappedIon2018,mehdiScalableQuantumComputation2020,ratcliffeMicromotionenhancedFastEntangling2020,mehdiFastMixedspeciesQuantum2025}. These SDK pairs were assumed to be separated by a very short delay such that free evolution of the ion motion between them can be ignored i.e. $\hat{U}_{\rm{SDK, pair}}^{(m)} = \hat{U}^{(m)}_{\rm{SDK}}(-\kappa)\hat{U}_{\rm{SDK}}^{(m)}(\kappa)=e^{2ik\hat{x}^{(m)}}\hat{\sigma}_{z}$. The distinction between this unitary and the unpaired SDK case (Eq.~\eqref{eq:SDKUnitary_PositionBasis}) is that the qubit spin-state is now conserved after each SDK, and the strength of each momentum transfer is doubled ($\pm 2\hbar k$) as shown in Fig.~\ref{fig:concept_diagrams1}(c). Additionally, there is no longer any dependence on the laser phase, $\phi$. 

Although fast gates using paired SDKs have been the focus of previous theoretical work, they are challenging to implement experimentally. The precise optical control required to create the short delay between SDKs in the pair to prevent motional evolution has limited demonstrations to gates using unpaired SDKs \cite{wong-camposDemonstrationTwoAtomEntanglement2017}. Furthermore, as each paired SDK requires double the number of pulses, imperfections in each pulse will double the total SDK infidelity. 

Figure~\ref{fig:unpaired_plot} compares the gate error for gate schemes optimized using paired and unpaired SDKs. We machine design fast gate schemes based on paired SDKs using the process outlined in \S~\ref{sec:gate_design}, however, as the strength of each SDK is now increased by a factor of $2$, the phase accumulation and motional restoration conditions (Eq.~\eqref{eq:entangling_phase_condition} and Eq.~\eqref{eq:motional_displacement_condition_APG}) are scaled accordingly: $\Theta_{\rm{paired}} = 4 \Theta_{\rm{unpaired}}$ and $\Delta \beta_{\alpha, \rm{paired}} = 2\Delta \beta_{\alpha, \rm{unpaired}}$ \cite{galeOptimizedFastGates2020}. Additionally, given gate schemes based on paired SDKs double the number of $\pi$-pulses used in the gate, we define $\mathcal{N}_{\rm{paired}} = 2\mathcal{N}_{\rm{unpaired}}$ and $f_{\rm{rep, paired}} = \frac{1}{2}f_{\rm{rep, unpaired}}$ to fairly compare the schemes. 

In the subsonic regime, the unpaired SDK schemes universally outperform the paired schemes with high-fidelity ($> 99.9\%$) solutions available with fewer than $50$ $\pi$-pulses. In this regime, the ion motion is no longer localized to the target ions, making the motional restoration needed to disentangle spin
motional degrees of freedom more difficult at the end of the gate. Using the unpaired scheme effectively halves the strength of each SDK to provide more targeted control of the ion motion, allowing entirely new gate solutions to be found that satisfy the motional restoration conditions (Eq.~\eqref{eq:motional_displacement_condition_APG}). 

For gate times significantly faster than the ion motion, the stronger momentum kicks provided by paired SDKs are advantageous as they enable the execution of the necessary phase-space trajectories on shorter timescales. However, the larger number of SDKs required on these timescales means that the total gate error will be dominated by SDK infidelities, rendering these benefits negligible.

\subsection{SDK repetition rate}
\begin{figure*}
	\centering
	\includegraphics[width=0.9\textwidth]{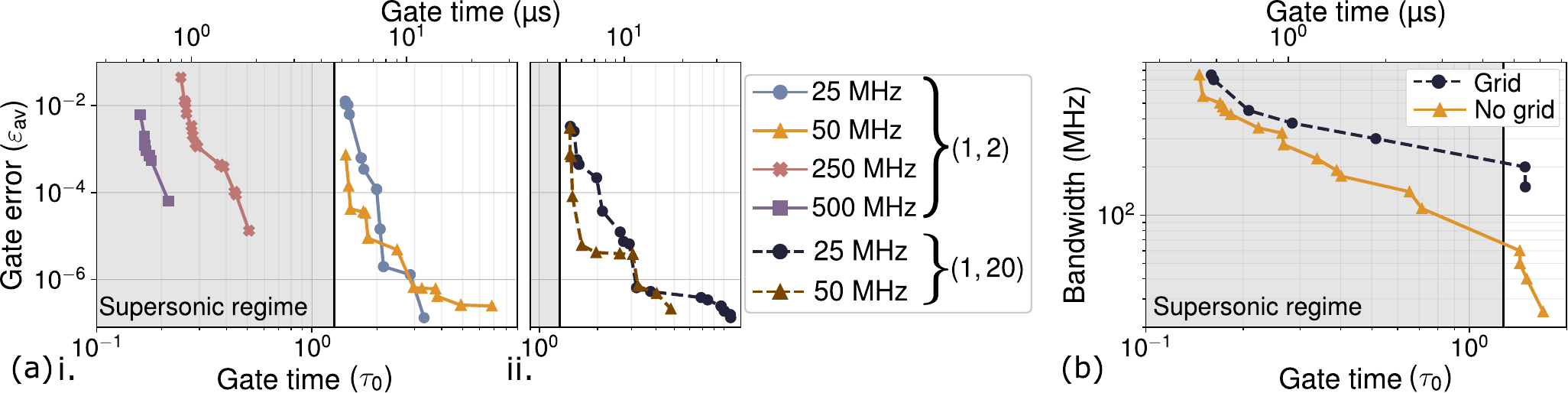}
	\caption{\textbf{Gate performance with different SDK repetition rates:} (a) Minimum SDK repetition rate required to perform gates above a $99.9\%$ threshold fidelity as a function of gate time. All gates were optimized for a local ion pair at the edge of a $20$-ion chain. Gate performance is also assessed when the SDK timings are fixed to a discrete grid that is defined by integer multiples of the SDK repetition rate. (b) Gate error (in the absence of pulse errors) as a function of gate time is shown for different SDK repetition rates for gates between local (b.i) and non-local (b.ii) ion pairings. The supersonic regime is indicated by the shaded region. All gate solutions assume a $20$-ion chain in an anharmonic trap characterized by $\kappa_{2} = 2.78\times 10^{-13}$ \unit{\joule\per\meter\squared} and $\kappa_{4} = 4.27\times 10^{-4}$ \unit{\joule\per\meter\tothe{4}}.}
	\label{fig:frep_plot}
\end{figure*}

In our gate searches, the SDK repetition rate is enforced by applying the constraint  $|t_{k} - t_{k+1}| \geq 1/f_{\rm{rep}}$ to the SDK sequences. Figure ~\ref{fig:frep_plot}(a) presents the trade-off between the gate error (before inclusion of SDK errors) and gate-time for solutions optimized with different minimum SDK repetition rates in a $20$ ion chain. To implement sub-microsecond gates with high fidelities, SDK repetition rates of over \SI{100}{\mega\hertz} are required, which is consistent with the results presented in Ref.\cite{mehdiFastEntanglingGates2021}. However, we find that extending our search to the subsonic regime enables high fidelity ($\varepsilon_{\rm{av}} \leq 10^{-4}$) gates between both local and non-local ions with with SDK repetition rates as low as \SI{25}{\mega\hertz} (i.e more than \SI{40}{\nano\second} between SDKs). 

In Fig.~\ref{fig:frep_plot}(b) we compare our results with optimized solutions where the SDK timings were further constrained to a discrete grid defined by integer multiples of the repetition rate: $t_{k} = n_{k}/f_{\rm{rep}}$ where $n_{k}$ is an integer \cite{galeOptimizedFastGates2020, mehdiFastEntanglingGates2021}. The inclusion of the grid constraint in our optimizations excludes the higher-fidelity solutions that exist at interim timings, which is particularly important when designing schemes with very low SDK repetition rates. As a result, we find gates with fidelities exceeding $99.9\%$ are infeasible for SDK repetition rates lower than \SI{150}{\mega\hertz} when this grid was enforced.

Enforcing the SDKs to arrive on this grid is advantageous in fast gate implementations where the SDK timings are limited by the repetition rate of pulsed lasers, where engineering non-uniform pulse timings requires complex optics such as nested delay lines \cite{bentleyFastGatesIon2013, wong-camposDemonstrationTwoAtomEntanglement2017}. However, given current laser repetition rates are on the order of hundreds of MHz to GHz \cite{hussainUltrafastHighRepetition2016, heinrichUltrafastCoherentExcitation2019}, enforcing a grid at lower repetition rates is unnecessary. In these cases, SDK timings within each pulse group are limited by other experimental factors such as the optical switching timescales \cite{johnsonUltrafastCreationLarge2017, heinrichUltrafastCoherentExcitation2019, wong-camposDemonstrationTwoAtomEntanglement2017} or the length of the pulse trains required to implement pulsed Raman SDKs \cite{mizrahiQuantumControlQubits2014, mizrahiUltrafastSpinMotionEntanglement2013, johnsonUltrafastCreationLarge2017, wong-camposDemonstrationTwoAtomEntanglement2017} (see \S~\ref{sec:SDK_implementation}).  We find enforcing a \SI{1}{\giga\hertz} grid post-optimization on the solutions in the subsonic regime of Fig.~\ref{fig:frep_plot}(a) introduces errors on the order of $10^{-4}$ to the total gate error. Alternatively, SDK implementations using modulated continuous wave lasers \cite{anProgrammableAdiabaticRapid2025, liuHighFidelityRamanSpinDependent2025} would enable even greater timing control, limited only by modulator bandwidths which are on the order of 10s of GHz \cite{anProgrammableAdiabaticRapid2025}.

\subsection{Dependence on system parameters}\label{subsec:eta_dependence}
\begin{figure*}
	\centering
	\includegraphics[width=\textwidth]{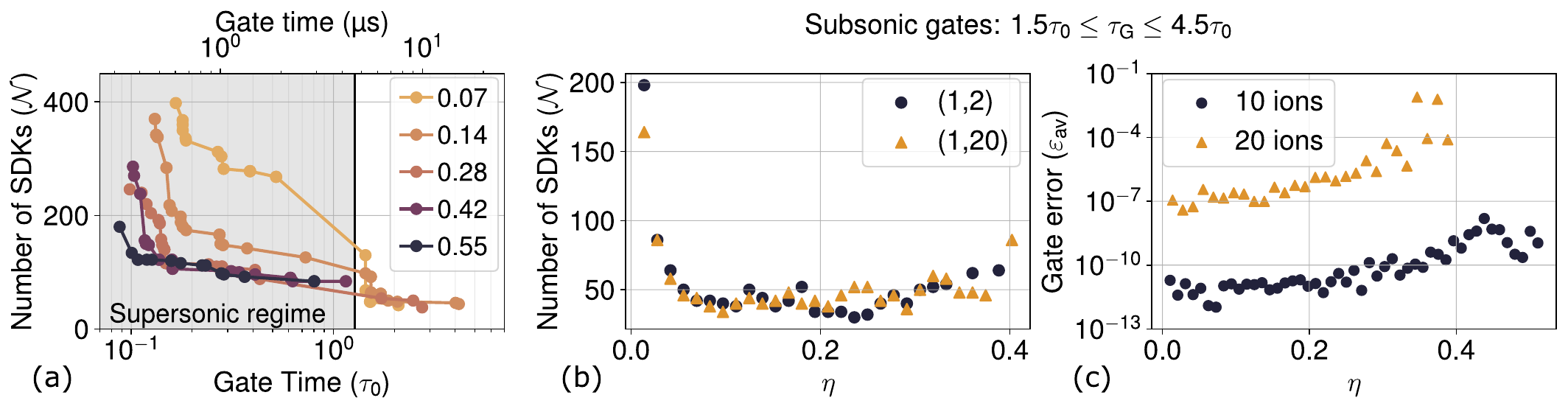}
	\caption{\textbf{Gate performance with different kick strengths in the absence of pulse errors:} (a) Number of SDKs required to perform a gate with greater than $99.9\%$ fidelity with varying momentum kick strengths ($\eta$) as a function of gate time. All gates were optimized for $2$ neighboring ions at the edge of a $20$-ion chain. The supersonic regime is indicated by the shaded region. (b-c) show gate performance in the subsonic regime for gate times $1.5\tau_{0}\leq \tau_{G} \leq 4.5\tau_{0}$. (b) Number of SDKs required to achieve a target fidelity greater than $99.9\%$ for both local and non-local ion pairings in a $20$-ion chain as a function of $\eta$. (c) Gate error (assuming perfect SDKs) as a function of $\eta$ for gates optimized between neighboring ions at the edge of $10$ and $20$-ion chains, constrained to fewer than $250$ SDKs. All gate solutions assume an infinite repetition rate (i.e. all SDKs in a group arrive simultaneously) and the minimum ion separation in the anharmonic trap was kept constant at \SI{3}{\micro\meter}.}
	\label{fig:eta_plot}
\end{figure*} 
The fast gate mechanism is compatible with a broad range of ion species, with the only ion-dependent contribution arising in the Lamb-Dicke parameter, $\eta$. In Figure~\ref{fig:eta_plot} we investigate the dependence of fast gate performance on $\eta$. For simplicity, we consider the ideal case where all the SDKs in a group arrive simultaneously, so there is no dependence on the SDK repetition rate.

In Fig.~\ref{fig:eta_plot}(a), we present the number of SDKs required to perform a gate above a threshold fidelity of $99.9\%$ as a function of gate time. In the supersonic regime, our results are consistent with previous work \cite{bentleyTrappedIonScaling2015}, where gates with larger momentum kicks require fewer SDKs. On these timescales, only the target ions are perturbed during the gate due to the separation of timescales between the gate dynamics and the collective trap motion. This simplifies the motional restoration needed to disentangle spin-motional degrees of freedom, however large phase-space displacements are required in order to accumulate the required phase on shorter timescales. Therefore, using a larger Lamb-Dicke parameter (e.g. by using a lighter ion such as $^{40}$Ca$^+$) enables the reduction in number of SDKs, which otherwise need to be concatenated with picosecond delays to achieve similar displacements from multi-photon recoils~\cite{bentleyFastGatesIon2013}.

In the subsonic regime, using larger kick strengths does not improve gate performance. Instead, in Fig.~\ref{fig:eta_plot}(a), the number of SDKs in high-fidelity solutions converges to $\mathcal{N} \approx 50$. Furthermore, we are no longer able to find high-fidelity solutions for larger Lamb-Dicke parameters. In subsonic gates, there is sufficient time to accumulate the required phase with a small number of kicks, so larger phase-space displacements during the gate are unnecessary however, the motion of all the ions becomes perturbed and thus gate fidelity is limited by motional restoration. In this case, a smaller Lamb-Dicke parameter enables fine-tuning of the state-dependent motional trajectories, as compared to using fewer SDKs with larger displacements in phase space. This becomes increasingly important in larger ion systems due to the increased number of motional modes that must be restored (see Fig.~\ref{fig:eta_plot}(c)).

However, the advantages of fine-tuning the motional trajectories provided by smaller kicks do not extend to arbitrarily small $\eta$ when stronger constraints are placed on the number of SDKs in the gate. For very small $\eta$, subsonic gates again become limited by the rate of phase accumulation, so more SDKs are required to achieve the necessary phase-space displacements. In Fig.~\ref{fig:eta_plot}(b), we find the number of SDKs required in high-fidelity gate solutions increases for $\eta < 0.1$.

The results in Fig.~\ref{fig:eta_plot}(b) suggest Lamb-Dicke parameters in the range $0.1-0.3$ are optimal for chains of $20$ ions. Although the Lamb–Dicke parameter is governed by ion-specific properties, its value can be controlled experimentally to fall within this range by adjusting the angle of the Raman beams used to drive SDKs. Specifically, moving the beams closer to the transverse axis reduces the wavevector coupling to the axial modes (see \S~\ref{sec:SDK_implementation}). For example, in the results presented in this work, we assume $^{133}$Ba as the candidate ion species with Raman beams angled $30^{\circ}$ from the transverse axis driving the SDKs, which yields $\eta = 0.14$ in $20$-ion chains. In contrast, a lighter ion species such as $^{40}$Ca would require a beam angle of $11^{\circ}$ to obtain the same Lamb-Dicke parameter (assuming the Raman beam wavelength is $393$nm \cite{heinrichUltrafastCoherentExcitation2019}).

\subsection{Ion scaling}
\begin{figure}
	\centering
	\includegraphics[width=0.48\textwidth]{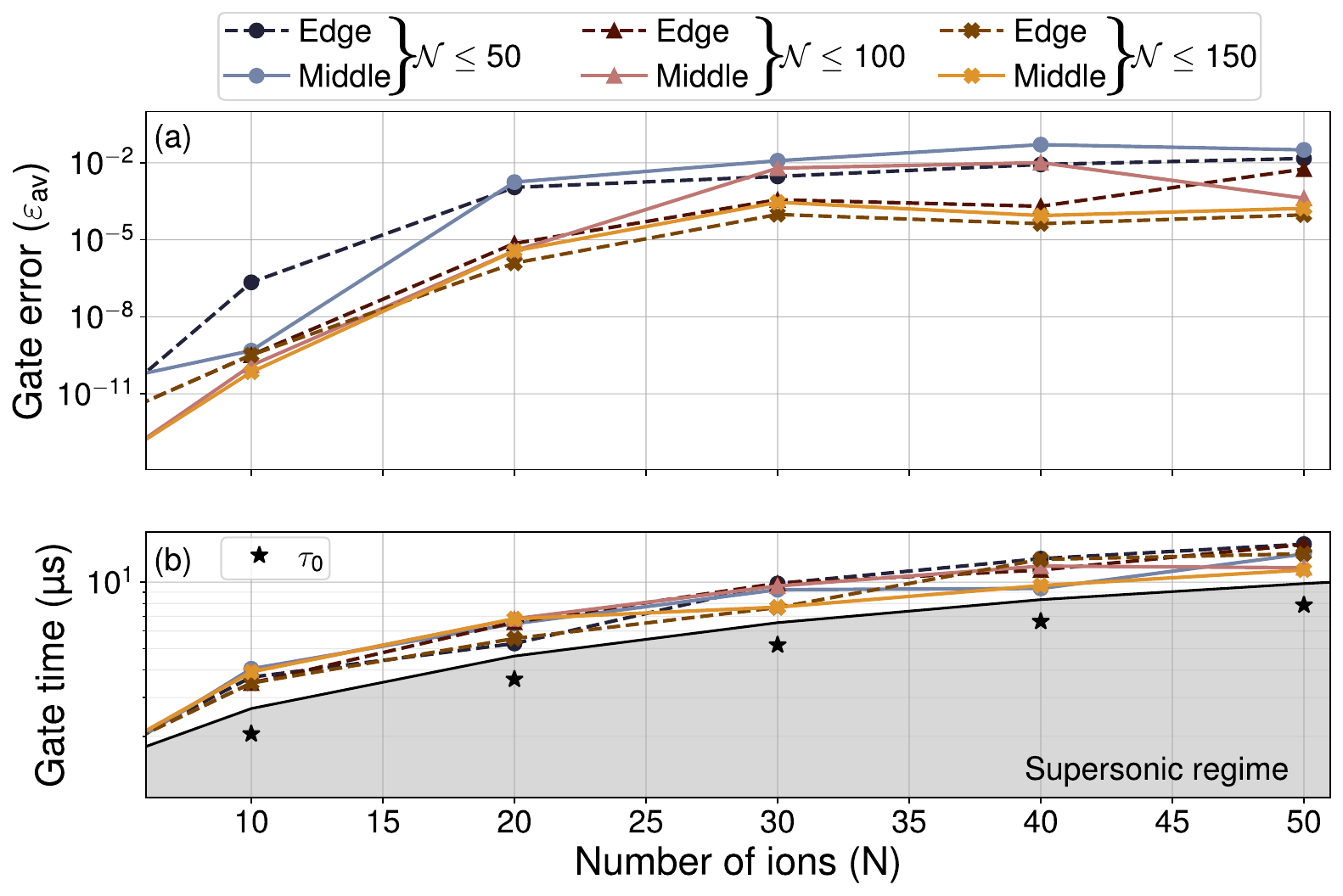}
	\caption{\textbf{ Theoretical gate error limit in long ion chains in the absence of pulse errors:} (a) Gate error (assuming perfect SDKs) and (b) gate time of optimized solutions in chains with varying numbers of ions. We consider gates between neighboring ions at the edge and in the middle of the chain with an upper-bound on the number of SDKs used in each gate. All gate times are in the subsonic regime between $1.4 - 2$ centre-of-mass oscillation periods. The supersonic regime is indicated by the shaded region. All gate solutions assume a \SI{500}{\mega\hertz} SDK repetition rate, and the minimum ion separation in the anharmonic trap was kept constant at \SI{3}{\micro\meter}.}
	\label{fig:scaling_plots}
\end{figure}
Our results suggest subsonic fast gates can achieve $99.9\%$ fidelity with less than 100 pulses, making these timescales highly favorable for near-term fast gate realization. We extend our analysis in Figure~\ref{fig:scaling_plots} to consider the scaling behavior of fast gates in this regime for ion chains of up to 50 ions. We exclusively consider local gates between nearest-neighbor ion pairs at the edge and in the center of the ion chain; however in the companion paper to this work \cite{savill-brownHighspeedHighconnectivityTwoqubit2025} we generalize this to consider the performance of non-local gates. For chains with more than $20$ ions the gate error plateaus which can be explained by the lower order motional modes dominating the interactions during the gates, which makes motional restoration easier \cite{savill-brownHighspeedHighconnectivityTwoqubit2025}. We find $99\%$ ($99.95\%$) fidelities are achievable (before the inclusion of SDK errors) for both nearest neighbor pairings using fewer than $50$ ($150$) SDKs and subsonic gate times below $2\tau_{0}$. For slightly longer gate times (up to $10\tau_{0}$), $99.95\%$ fidelities can be obtained with fewer than $50$ SDKs \cite{savill-brownHighspeedHighconnectivityTwoqubit2025}. 

As the number of ions was increased, higher-dimensional searches (i.e. increasing the number of potential SDK groups) were required to find high-fidelity gate solutions. While this was not associated with an increase in the total number of pulses, it suggests the timing freedoms provided by these higher-dimensional searches become increasingly necessary to control the larger number of motional modes. We find optimization of high-fidelity subsonic gates in chains of more than $50$ ions is infeasible with the constraints we place on our optimizations. This challenge does not arise in the supersonic regime, where distant ions are unaffected by the localized fast gate dynamics \cite{mehdiFastEntanglingGates2021}. Therefore, supersonic local gates can be performed in arbitrarily long chains \cite{mehdiFastEntanglingGates2021}, at the cost of losing connectivity between non-local ion pairs \cite{savill-brownHighspeedHighconnectivityTwoqubit2025} and increased SDK requirements.  

\begin{figure*}
	\centering
	\includegraphics[width=\textwidth]{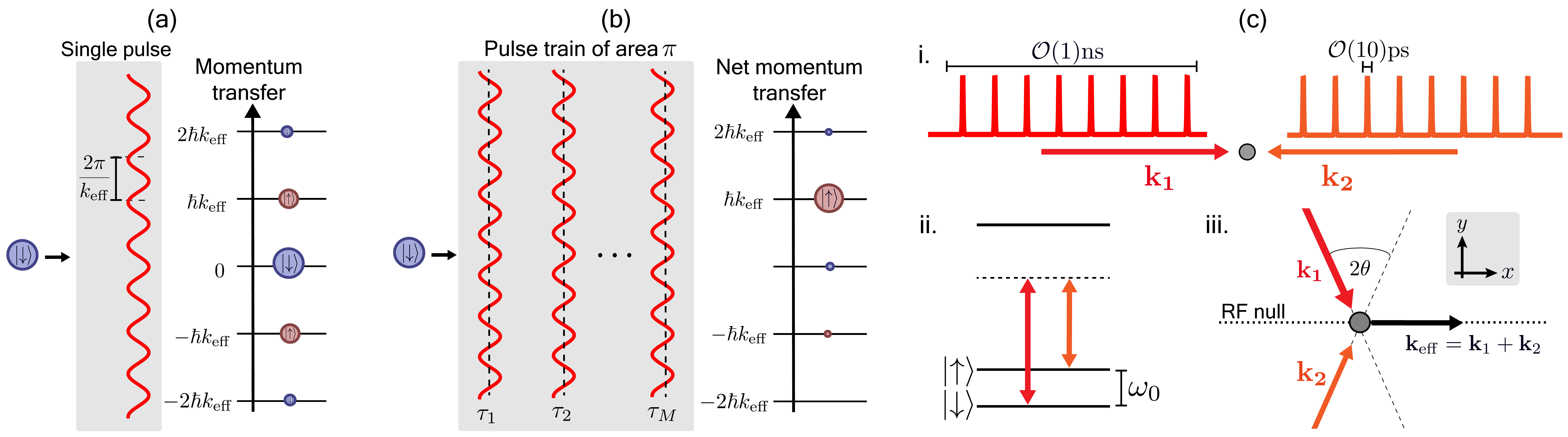}
	\caption{\textbf{Two-photon SDK scheme for hyperfine qubits using stimulated Raman transitions.} (a) Illustration of Kaptiza-Dirac scattering of an initial atomic wavepacket in the $\ket{\downarrow}$ spin-state (blue) by the optical lattice formed by the two Raman beams (red), for a single resonant pulse. The translational invariance of the lattice results in diffraction of the wavepacket into quantized momentum orders, where the $n$-th diffraction order is correlated with $n$ spin-flips and its amplitude weighted by an $n$-th order Bessel function of the first kind (see Eq.~\eqref{eq:BroadBandPulse_Unitary}). (b) High-fidelity SDKs may be realized using a broadband pulse train with total pulse area $\pi$ and numerically optimized timings~\cite{mizrahi_ultrafast_2013} such that population coherently accumulates in the target momentum order. (c) Schematic of laser geometry for implementing the SDK scheme illustrated in (b). i. The ion is addressed by a nanosecond sequence of $M\lesssim$ \SI{10}{\pico\second} pulses from each of the Raman beams with wavevectors $\bm{k}_{1,2}$. ii. Level diagram for two hyperfine-split ground states, $\ket{\uparrow}$ and $\ket{\downarrow}$, coupled via a two-photon transition detuned from a highly excited state (dashed line). Resonant two-photon transitions are achieved by matching the frequency separation between the two Raman beams to the qubit frequency $\omega_0 \sim$ \SI{10}{\giga\hertz}. iii. Beam geometry for coupling tilted transverse Raman beams to the axial motion of the ion -- i.e. the effective wavevector $\bm{k}_{\rm eff}$ is aligned with the null of the radio-frequency (RF) trap. \label{fig:RamanBeamGeometry}}  
\end{figure*} 
\section{State-dependent kick implementation}\label{sec:SDK_implementation}
As stated earlier, the theoretical framework of fast two-qubit gates is agnostic to the exact implementation of the impulsive SDKs, and it is thus compatible with a range of methodologies that have been experimentally demonstrated in trapped-ion systems. This includes single-photon approaches that implement SDKs with single resonant pulses at the repetition rate of the laser~\cite{heinrichUltrafastCoherentExcitation2019,hussainUltrafastHighRepetition2016,guoPicosecondIonqubitManipulation2022}, and two-photon Raman SDK schemes suitable for coupling hyperfine or Zeeman split qubit states~\cite{campbellUltrafastGatesSingle2010,mizrahi_ultrafast_2013,mizrahiQuantumControlQubits2014,johnsonUltrafastCreationLarge2017,wong-camposDemonstrationTwoAtomEntanglement2017}.

\subsection{Raman SDKs for hyperfine qubits}
Here we focus on Raman SDK schemes that are suitable for hyperfine qubits typically used for quantum computation due to their qubit lifetimes, such as $^{171}$Yb or $^{133}$Ba. The key challenge to such schemes is multi-photon transitions into unwanted momentum states due to the diffraction of the atomic wavepacket from the optical lattice formed by the two Raman beams~\cite{mizrahi_ultrafast_2013,mizrahiQuantumControlQubits2014} (the Kapitza-Dirac effect~\cite{Gould1986}), illustrated in Figure \ref{fig:RamanBeamGeometry}. To understand this, we can consider an approximate Hamiltonian describing the atom-light interaction with a single ion assuming two-photon resonance and that the pulse duration is sufficiently short that motional evolution may be neglected: $\hat{H}_I\approx \Omega(t)\cos(k_{\rm eff}\hat{x}+\phi)\hat{\sigma}_x $, where $k_{\rm eff}$ is the effective wavevector of the two-photon transition, shown in Figure~\ref{fig:RamanBeamGeometry}. Using $k_{\rm eff}\hat{x} = \eta(\hat{a}+\hat{a}^\dag)$ and assuming a pulse with total pulse area $\theta = \int \Omega(t)dt$ and duration much faster then the hyperfine qubit splitting, $t \ll \frac{2\pi}{\omega_{0}}$, the following expression can be derived for the unitary of a single pulse~\cite{mizrahi_ultrafast_2013,mizrahiQuantumControlQubits2014}:
\begin{align}
\label{eq:BroadBandPulse_Unitary}
    \hat{U}(\theta) = \sum_{n=-\infty}^\infty i^n J_n(\theta) e^{in\phi} \hat{D}(in\eta)\hat{\sigma}_x^n \,.
\end{align}
This expression is notably distinct from the unitary for an ideal SDK, Eq.~\eqref{eq:SDKUnitary_ModeBasis}, as it includes off-resonant multi-photon transitions ($n\neq 1$) that deliver multiple photon recoils and spin-flips with amplitude weighted by the $n$-th order Bessel function $J_n(\theta)$ -- see Fig.~\ref{fig:RamanBeamGeometry}(b).  This precludes the realization of high-fidelity SDKs using this scheme with a single pulse, as unwanted momentum orders will have non-negligible population~\footnote{It is possible to realize a single-pulse Raman SDK between Zeeman-split qubit states using polarization selective transitions~\cite{putnamImpulsiveSpinmotionEntanglement2024}, however this scheme requires a large frequency separation between the pulse bandwidth and qubit splitting that is generally not achievable for hyperfine qubits.}.

High-fidelity Raman SDKs can be realized by using a sequence of $M$ pulses, each with pulse area $\pi/M$ as demonstrated in Refs.~\cite{mizrahi_ultrafast_2013,mizrahiQuantumControlQubits2014} and illustrated in Fig.~\ref{fig:RamanBeamGeometry}. Numerical optimization of the delays between pulses in the sequence ensures that the population accumulates coherently in the targeted motional state, with destructive interference of off-resonant processes into unwanted momentum. This allows for SDK fidelities above $99.9\%$ with $M<10$ pulses with a total duration of a few nanoseconds~\cite{mizrahi_ultrafast_2013}, with further improvements for longer pulse sequences~\cite{mizrahiQuantumControlQubits2014}. In principle, optimal control of the pulse amplitudes and phases could enable much higher SDK fidelities.

In the experimental demonstrations of Refs.~\cite{mizrahi_ultrafast_2013,johnsonUltrafastCreationLarge2017,wong-camposDemonstrationTwoAtomEntanglement2017}, the effective SDK duration (duration of the pulse sequence) is approximately \SI{2.5}{\nano\second}, which corresponds to an SDK repetition rate of \SI{400}{\mega\hertz}. Even stricter constraints on the minimum timing between SDKs arise from the need to switch the direction of successive kicks, e.g. by rotating the polarization of the Raman beams to swap the directions of the counter-propagating beam (see Fig.~\ref{fig:RamanBeamGeometry}(c))~\cite{johnsonUltrafastCreationLarge2017,wong-camposDemonstrationTwoAtomEntanglement2017}. For example, the experiment of Ref.~\cite{johnsonUltrafastCreationLarge2017} used a Pockell cell to achieve pulse direction swapping in approximately \SI{12}{\nano\second}, which restricts the effective SDK repetition rate to $f_{\rm rep}\lesssim$ \SI{83}{\mega\hertz}. These limitations to the effective SDK repetition rate are not fundamental to the Raman SDK scheme and can be bypassed using optical delay lines \cite{bentleyFastGatesIon2013}. However, this is not necessary as high-fidelity two-qubit gates between arbitrary qubit pairs can still be achieved in the subsonic regime even with SDK repetition rates as low as \SI{25}{\mega\hertz}, as shown in Fig~\ref{fig:frep_plot} of \S~\ref{sec:results}.

Recent theoretical work has also proposed an alternative approach to mitigate the lattice-induced diffraction in Raman SDKs by instead amplitude-modulated continuous-wave (CW) Raman beams rather than subdividing each SDK into an $M$-pulse train \cite{anProgrammableAdiabaticRapid2025, liuHighFidelityRamanSpinDependent2025}. In these schemes, the SDK is generated by shaping the temporal envelope of the Raman coupling to engineer an effective impulsive interaction while suppressing residual errors through optimal control of the modulation waveform. In this framework, the relevant experimental constraints are set by the modulation bandwidth of electro-optic or acousto-optic modulators, which can reach tens of GHz, rather than by the repetition rate of a mode-locked laser.

\subsection{Raman beam geometry addressing the axial modes}
In the fast gate schemes considered in this work, we consider SDKs on individually addressed ions that couple to the axial motional modes of a quasi-uniform ion chain. Both aspects pose a challenge for long ion chains and trapping geometries with limited optical access. 

We propose that axial motion can be combined with individual addressing by using a pair of tightly-focused Raman beams that are equally tilted from the RF null of the trap~\cite{putnamImpulsiveSpinmotionEntanglement2024}, as shown in Fig.~\ref{fig:RamanBeamGeometry}(c). If we assume the wavevectors of the two beams have equal magnitude -- i.e. $k\equiv |\bm{k}_1|=|\bm{k}_2|$ --  then the effective wavevector coupling to the axial modes of the trap is given by:
\begin{align}
    k_{\rm eff} = 2k\sin(\theta) \,,
\end{align}
where $\theta$ is the angle of the beams as shown in Figure \ref{fig:RamanBeamGeometry}(c). In order for this approach to be compatible with long ion chains, the beam angle needs to be relatively small in order to prevent cross-talk with neighboring ions -- though this can be compensated by increasing the interion distance at the cost of a reduced axial trapping frequency (and by extension, increased gate times). For all the results presented in this work, we take the angle between the two beams to be $2\theta = 60^\circ$ such that the magnitude of the kick is equivalent to a single-photon transition, i.e. $k_{\rm eff}=k$.  If the counter-propagating Raman beams are not equally tilted from the RF null of the trap there will be excitations to the radial motional modes of the ion chain, in addition to the axial modes. Specifically, for small beam angle misalignments, $\delta \theta \ll1$, we expect radial excitations $\eta_{r} \sim \eta\delta\theta$, which will contribute to the total gate infidelity with $\sim \eta_{r}^{2}$, meaning the gate will be insensitive to first-order angle errors.Future work will explore fast gate performance with off-axis SDKs that couple to both the radial and axial motion of the crystal.

\section{Error analysis}\label{sec:robustness}

\begin{figure*}
	\centering
	\includegraphics[width=0.9\textwidth]{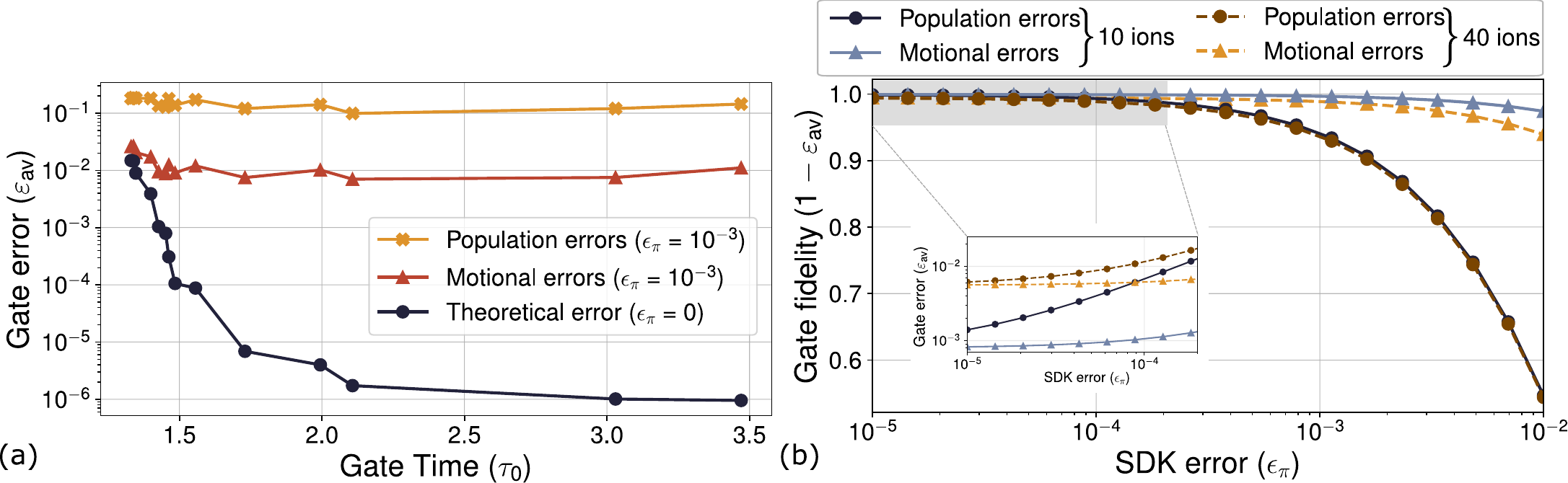}
	\caption{\textbf{Fast gate performance in the presence of SDK errors:} (a) Total gate error in the presence of population inversion errors and motional state diffusion errors arising from imperfect SDKs for $\epsilon_{\pi} =10^{-3}$ as a function of gate time. All gates were optimized for a pair of ions at the edge of a $20$-ion chain with fewer than $100$ SDKs, assuming a \SI{500}{\mega\hertz} SDK repetition rate. The theoretical gate error assuming perfect SDKs ($\epsilon_{\pi} = 0$) is also shown. (b) Impact of SDK errors on the fidelity for exemplary subsonic gates between a pair of ions at the edge of a $10$-ion chain (solid lines) and a $40$-ion chain (dashed lines). The inset plot shows the gate error in the shaded region. Both gates were composed of $30$ SDKs and had ideal ($\epsilon_{\pi} = 0$) fidelities greater than $99\%$. Population inversion errors are assumed to degrade the gate fidelity according to $(1-\epsilon_{\pi})^{2\mathcal{N}}$, while motional diffusion errors are simulated using a Monte-Carlo model with $10^{4}$ sample sets.}
	\label{fig:SDK_errors}
\end{figure*} 

\subsection{SDK diffraction errors}

The unitary for an imperfect SDK is  \cite{galeOptimizedFastGates2020} \begin{equation}
\label{eq:SDK_error_unitary}
    \hat{U}_{\rm{SDK}}(\theta) = \hat{\sigma}_{x}e^{i(k_{\rm{eff}}x + \phi)\hat{\sigma}_{z}}\cos\theta + \hat{\mathds{I}}\sin\theta~,   
\end{equation} where $\theta$ is the rotation error. This is a simplified model based on a single-photon transitions where $\epsilon_{\pi} = |\theta|^{2}$ is the transition error arising from an imperfect $\pi$-pulse, however, this can be generalized to include the higher order diffraction terms that arise in multi-photon transitions as discussed in \S~\ref{sec:SDK_implementation}. In the above expression, the identity term generates both an incorrect internal state and incorrect motional state while the $\hat{\sigma}_{x}$ operation generates the correct internal state but the motional state is incorrect. In the ideal case ($\theta = 0$), the above expression will reduce to the ideal SDK unitary given in Eq.~\eqref{eq:SDKUnitary_PositionBasis}.

As discussed in \S~\ref{sec:gate_design}, we use an upper-bound estimate on the actual gate error in the presence of imperfect SDKs in our optimizations. This is based on a worst-case approximation that assumes each SDK causes an imperfect $\pi$-rotation on the Bloch sphere, so there is no overlap between the erroneous states and the target state \cite{galeOptimizedFastGates2020}. Under this assumption, for a gate with $\mathcal{N}$ SDKs that each have error probability $\epsilon_\pi\ll 1$, the 2Q gate fidelity will degrade as $(1-\epsilon_{\pi})^{2\mathcal{N}} \approx 1-2\epsilon_{\pi}\mathcal{N}$.

In Figure~\ref{fig:SDK_errors}, we demonstrate the impact of these population inversion errors on the total gate fidelity. The total gate error in the presence of population inversion errors scales as $\mathcal{N}\epsilon_{\pi}$ (with $\mathcal{N}$ on the order of $10^{2}$ in this work), which is consistent with previous results \cite{mehdiFastEntanglingGates2021}. Therefore, achieving gate fidelities above  $99\%$ ($99.9\%$) requires population inversion errors to be suppressed to the $10^{-4}$($10^{-5}$) level. In principle, population inversion errors can be suppressed using robust quantum control such as STIRAP \cite{bergmannRoadmapSTIRAPApplications2019}, or composite pulse sequences \cite{wimperisBroadbandNarrowbandPassband1994, cumminsTacklingSystematicErrors2003}. Alternatively, if the pulse area errors between the two SDKs are strongly correlated, the simplest approach to suppress these errors is by engineering a $\pi$ phase shift between consecutive SDKs  \cite{taylorStudyFastGates2017}. 

In the case where population inversion errors are perfectly suppressed, an errant SDK will still result in errant spin-motional trajectories, which will dephase the two-qubit state. For simplicity, we assume that the dominant error channel is that there is no kick. Therefore, the overlap between the ideal state and the errant motional state of a single mode will be  $|\langle\beta|\beta+i\eta\rangle|^2 = e^{-\eta^2}\approx 1-\eta^2$ for $\eta\ll 1$. Because the error from imperfect spin-motional decoupling is proportional to the final motional state overlap, the 2Q gate error is expected to scale as $\eta^2 \epsilon_\pi \mathcal{N}$.

To simulate the effects of motional errors produced by imperfect SDKs, we use a Monte-Carlo technique that introduces errors probabilistically at each kick for a large ($10^{4}$) number of sample sets, as described in Appendix~\ref{sec:Monte-Carlo_Errors}. In a given SDK sequence, we take the probability of an erroneous kick to be $\epsilon_{\pi}$, which will result in either a kick in the wrong direction or no kick to both ions. This model will be generalized to account for Kapitza-Dirac scattering into higher-order momentum modes in a later manuscript. 

Fig.~\ref{fig:SDK_errors} shows the mean gate error as a function of different SDK error probabilities, $\epsilon_{\pi}$. In Fig.~\ref{fig:SDK_errors}(a)  we observe up to an order of magnitude improvement in the total gate error if imperfect SDKs only cause errors to the motional states. Fig.~\ref{fig:SDK_errors}(b) shows fidelities greater than $90\%$ are possible for $\epsilon_{\pi} = 10^{-2}$ in chains of up to $40$ ions, compared to fidelities less than $60\%$ expected in the presence of population inversion errors of the same magnitude. 

We also compare the achievable gate fidelities for gates optimized in chains with different numbers of ions in Fig.~\ref{fig:SDK_errors}(b). In longer ion chains, keeping the ion separation constant reduces the center-of-mass motional frequency, resulting in a larger Lamb-Dicke parameter; $\eta = 0.18$  in the $40$-ion chain compared to $\eta = 0.1$ in the $10$-ion chain. As the motional state errors scale with $\eta^{2}$, we observe more significant degradation in the fidelity of the gate in the $40$-ion chain. In Raman SDK implementations, the Lamb-Dicke parameter can be reduced by decreasing the Raman beam angle to the transverse axis (e.g. $\theta \approx 17^{\circ}$ would result in $\eta = 0.1$ for a $40$-ion chain). 
   
Current experiments achieve SDK fidelities around $99\%$ \cite{johnsonUltrafastCreationLarge2017, hussainUltrafastHighRepetition2016, johnsonSensingAtomicMotion2015} which is sufficient to perform gates with fidelities of $95\%$ in long ion chains for $\eta \sim 0.1$ provided population inversion errors can be suppressed. Furthermore, existing theoretical Raman SDK schemes suggest SDKs can be implemented with $99.99\%$ fidelities \cite{mizrahiQuantumControlQubits2014}, which would reduce 2Q gate errors to the $10^{-3}$ level.  Alternatively, recent proposals based on optimized CW SDK schemes suggest this error can be further improved, limited only by spontaneous emission at the $10^{-7}$ level \cite{liuHighFidelityRamanSpinDependent2025}

\subsection{Timing errors}
\begin{figure*}
	\centering
	\includegraphics[width=0.8\textwidth]{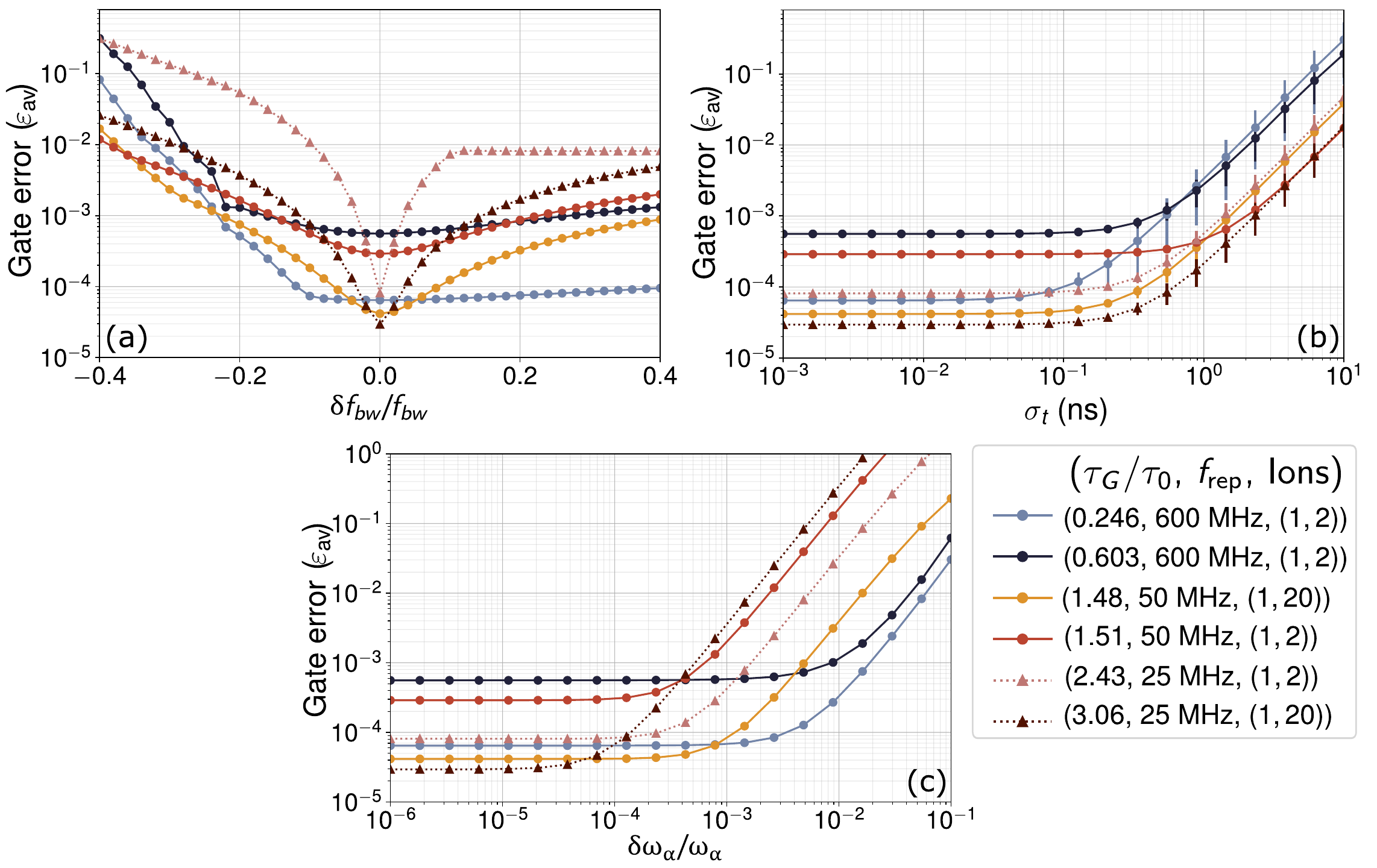}
	\caption{\textbf{Experimental robustness of fast gate schemes:} Impact of (a) shifts in the SDK repetition rate, (b) random jitters in the SDK group timings and (c) alterations to the mode spectrum on the infidelity of select local and non-local gates. (a) A static shift in the SDK repetition rate relative to the repetition rate used in the original optimization was applied to each pulse sequence. (b) Each pulse group was offset by sampling a Gaussian distribution centered around the originally optimized timings with a standard deviation $\sigma_{t}$. The mean infidelity of $10^{4}$ sample sets is plotted as a function of the distribution width, with the standard deviation indicated by the error bars. (c) A static shift was applied to each of the motional mode frequencies $\omega_{\alpha}$. All gate solutions assume a $20$-ion chain in an anharmonic trap characterized by $\kappa_{2} = 2.78\times 10^{-13}$ \unit{\joule\per\meter\squared} and $\kappa_{4} = 4.27\times 10^{-4}$ \unit{\joule\per\meter\tothe{4}}.}
	\label{fig:errors}
\end{figure*} 

\subsubsection{Repetition rate errors}
Miscalibrations of the SDK repetition rate typically arise in pulsed lasers due to drifts of the repetition rate. These drifts are much slower than the gates we consider \cite{hayesEntanglementAtomicQubits2010, mizrahiQuantumControlQubits2014, johnsonUltrafastCreationLarge2017}, so we treat them as a systematic shift in the SDK repetition rate during the gate. Figure~\ref{fig:errors}(a) demonstrates the impact of this shift on the gate error for a set of optimized gate solutions. We find fast gate schemes maintain fidelities above $99.9\%$ despite systematic shifts of up to $10\%$ of the original repetition rate.

\subsubsection{Pulse timing errors}
A separate source of error is the imperfect timing of SDKs, which we analyze independently of variations in the repetition rate. Given that pulsed lasers exhibit sub-picosecond timing jitters \cite{clarkPhaseNoiseMeasurements1999}, we assume the timings within each group to be highly stable as they are locked to the repetition rate of the laser. Therefore, the most likely source of error arises from incorrectly timing the first SDK in each group, which subsequently offsets the rest of the SDKs in that group. This also breaks the anti-symmetry built into our gate scheme, meaning momentum restoration is no longer guaranteed at the end of the gate. We model this error by adding a Gaussian noise distribution to the original timings of each pulse group. In Fig.~\ref{fig:errors}(b) we observe shifts of up to \SI{1}{\nano\second} introduce errors on the order of $10^{-4}$. Therefore, achieving fidelities of $99.9\%$ requires sub-nanosecond control over the SDK timings which is feasible for the pulse trains that are typically used to drive effective $\pi$-pulse transitions in hyperfine qubits \cite{mizrahiUltrafastSpinMotionEntanglement2013}.

\subsection{Stray electric fields}
In Fig.~\ref{fig:errors}(c) we investigate the impact of a common frequency shift to each of the motional modes, which alters the motional dynamics during the gate. As motional dephasing occurs on timescales on the order of \unit{\milli\second} \cite{talukdarImplicationsSurfaceNoise2016}, we treat the frequency shift as static within each gate. We find supersonic gates are stable despite frequency shifts on the order of $10^{-2}$. However, greater stability is required to perform subsonic gates, with frequency shifts on the order of $10^{-4}$ required to maintain $99.9\%$ fidelities, which can be achieved in existing traps \cite{charlesdoretControllingTrappingPotentials2012}.

\subsection{Hot motional states}
\begin{figure}
	\centering
	\includegraphics[width=0.45\textwidth]{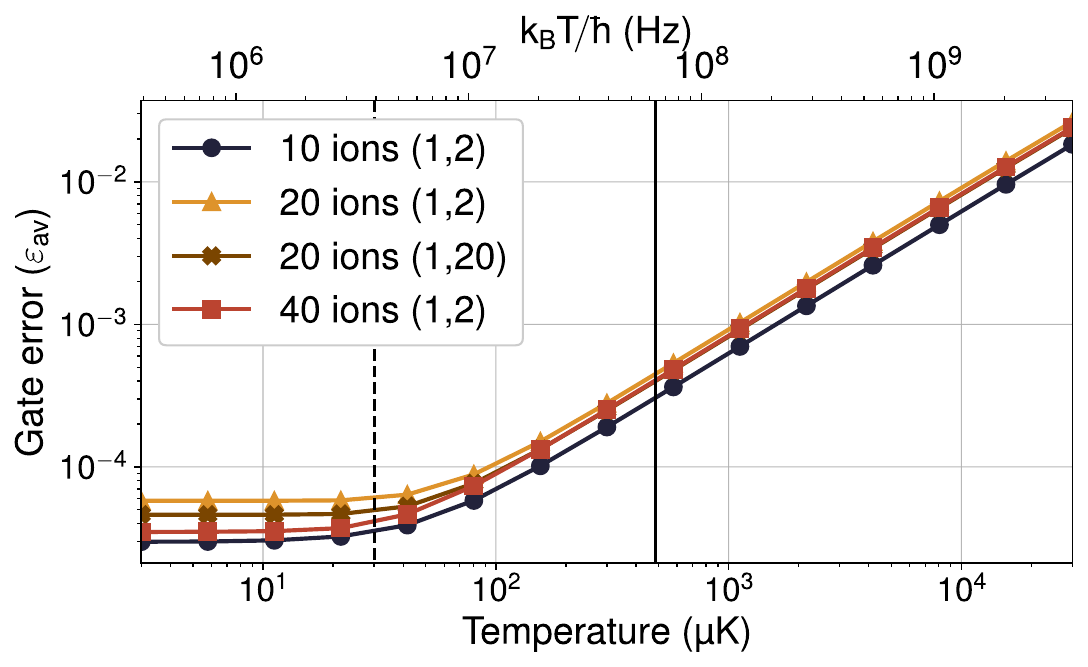}
	\caption{\textbf{Motional infidelity as a function of ion temperature:}  Contribution of residual spin-motional entanglement at the end of the gate to the total gate error is shown for exemplary local and non-local subsonic gates in different chain lengths as a function of the temperature of the ions. The dashed vertical line indicates the temperature we have assumed in our optimizations ($T =$ \SI{30}{\micro\kelvin}). The solid vertical line indicates the Doppler temperature ($T_{D} =$ \SI{486}{\micro\kelvin}). }
	\label{fig:temp_plot}
\end{figure} 

The relative phase acquired by different two-qubit states during a 2Q entangling operation is independent of the mode occupations. However, for higher mode occupations, the total gate error will increase when there is residual spin-motional entanglement at the end of the operation (Eq.~\eqref{eq:infidelity}). In Figure~\ref{fig:temp_plot} we show the motional error contribution to the total gate infidelity at different system temperatures for a set of optimal gate solutions in different chain lengths. We find cooling to the Doppler temperature $\Gamma\hbar/2k_{B} =$ \SI{486}{\micro\kelvin} ($\bar{n}_{\rm{COM}} = 12.3$ in a $10$-ion chain) enables gate errors less than $10^{-3}$ in chains with up to $40$ ions. Mode occupations below the Doppler cooling limit, such as those we consider in this work for a \SI{30}{\micro\kelvin} system temperature  ($\bar{n}_{\rm{COM}} = 0.85$ in a $10$-ion chain), further reduce the gate error by an order of magnitude. Sub-Doppler cooling of long-ion chains to such temperatures is achievable using fast cooling techniques based on electromagnetically-induced transparency~\cite{fengEfficientGroundStateCooling2020, wuElectromagneticallyinducedtransparencyCoolingTripod2025}. 

We emphasize that while the fast gate mechanism is insensitive to larger motional occupations (i.e. outside the Lamb-Dicke regime), single-qubit control becomes more challenging for high temperatures in long ion chains due to the thermal motion of ions within the beam waist of focused laser beams transverse to the trap~\cite{johnsonSensingAtomicMotion2015}. Practical quantum computation with the scheme proposed in this work may benefit from sympathetic cooling to mitigate this effect in long chains containing dozens of ions~\cite{cetinaControlTransverseMotion2022}.

\subsection{Single-qubit rotations due to the optical phase}
As we showed in \S~\ref{sec:fast_gate_theory}, the unitary operator for a fast gate operation  based on unpaired SDKs (Eq.~\eqref{eq:GateUnitary_PhaseGate}) has a term arising from the laser phase which describes a single-qubit rotation on each of the addressed ions. While this does not affect the 2Q entangling operation, noise in the laser phase will dephase the qubits. Due to the antisymmetric constraint on the fast gate solutions considered in this work, we can express the final single-qubit rotation (Eq.~\eqref{eq:laser_phase}) due to the optical phase of each SDK as a sum of relative phases. This can be shown explicitly by imposing the anti-symmetric constraint of the APG scheme ($z_{j+\mathcal{N}/2} = -z_{j}$ and $t_{j+\mathcal{N}/2}=t_j+\tau_{\rm G}/2$; see \S~\ref{sec:gate_design}) on Eq.~\eqref{eq:laser_phase}, which gives:
\begin{align}
    \Phi_{\rm{1Q}} 
    = \sum_{j = 1}^{\mathcal{N}/2}z_{j}\left[\phi(t_{j}) - \phi(t_{j} + \frac{\tau_{G}}{2})\right] \,.\label{eq:APG_laser_phase}
\end{align}

This illustrates that the single-qubit rotations during fast gates depend only on differential phases between SDKs separated by half the gate time, $\tau_{\rm G}/2$. The final single-qubit phase can be compensated for by single-qubit operations, or eliminated entirely at the end of the gate by choosing the phases of each SDK appropriately; e.g. by applying a $\pi$-phase shift to each of the SDKs in the second half of the gate.

In the Raman SDK scheme, the laser phase from each SDK can be expressed as~\cite{wong-camposDemonstrationTwoAtomEntanglement2017} $\phi(t) = \omega_A t + \phi_0$. In this expression, $\omega_A$ is the two-photon detuning of the Raman beams and $\phi_0$ is the differential phase between the two Raman beams, assumed to be constant over a single fast gate, but varying between operations. Phase noise in this scheme arises from the differential phase, $\phi_0$, which is difficult to stabilize experimentally \cite{wong-camposDemonstrationTwoAtomEntanglement2017}. From Eq.~\eqref{eq:APG_laser_phase}, fast gates using the APG scheme are entirely insensitive to this differential phase. 

\subsection{Additional errors}
The effects of other error sources have been analyzed in previous works, however, we list the key points here for convenience. 

\subsubsection{Motional heating}
This was previously investigated in Ref.~\cite{taylorStudyFastGates2017}, where the authors found gate fidelities were significantly damaged if heating of the axial modes occurs during the operation. The gates considered in this work are much faster than trap heating rates~\cite{bruzewiczMeasurementIonMotional2015, hartyHighfidelityPreparationGates2014, paganoCryogenicTrappedIonSystem2018}, making them resilient to this type of error.  Specifically, for heating rates of $1-100$ quanta/s \cite{bruzewiczMeasurementIonMotional2015,niedermayrCryogenicSurfaceIon2014, schaferFastQuantumLogic2018}, we expect 2Q gate errors ($\varepsilon \sim \tau_{G}\dot{\bar{n}}$) on the order of $10^{-5}-10^{-3}$ to be introduced \cite{mehdiFastMixedspeciesQuantum2025, taylorStudyFastGates2017}. 

\subsubsection{Coulomb nonlinearities}
For efficient computation of the mode structure in long ion chains, we truncate the effective potential experienced by each ion to second order (see Appendix~\ref{sec:ode_appendix}). Ref.\cite{galeOptimizedFastGates2020} found that non-linearities in the Coulomb potential introduce errors on the order of $10^{-6}$. We anticipate that errors from the Coulomb non-linearity will be smaller than previously reported, because the gates in this work utilize fewer pulses and cover a lower momentum range.

\subsubsection{Anharmonicities in the trapping potential}
We also consider anharmonicities in the trapping potential due to the quartic term. On length scales smaller than the ion separation, these anharmonicities are small relative to the harmonic potential ($z_{\rm{min}} \ll \sqrt{\kappa_{2}/\kappa_{4}}$). Given that the ions only experience small displacements relative to their separation ($\sim 0.1\%$), the anharmonicity only contributes a shift in the normal mode frequencies and couplings \cite{homeNormalModesTrapped2011}. This shift was included in our calculations of the mode structure.

\section{Conclusions}

In this work, we have developed a refined theoretical model for designing fast gate protocols for implementation in quasi-uniform linear ion chains that are compatible with near-term laser control capabilities. We introduced a new approach for optimizing sequences of SDKs that incorporates experimental challenges with SDK implementation while also providing greater control over the motion in larger-ion crystals. These new protocols outperform previous gate schemes in large-ion systems when stronger constraints are placed on the number and timing of the SDKs. While SDK errors remain the primary technological limitation to fast gate implementation, we identify the subsonic regime to be highly favourable for near-term experimental implementation of fast gates in long ion chains due to the reduced SDK requirements on these timescales. We have also investigated the robustness of select gate schemes against other experimental errors and our results indicate high fidelities can be achieved within existing technological constraints. 

The data that support the findings of this article is openly available \cite{savill-brownDatasetErrorResilientFast2025}.
\section{Acknowledgements}
This research was undertaken with the assistance of supercomputing resources and services from the National Computational Infrastructure, which is supported by the Australian Government. ISB thanks the support provided by the D.N.F Dunbar Honours Scholarship and the Australian Government Research Training Program (RTP) Scholarship.

\appendix

\section{Anharmonic trapping potential parameters}\label{sec:trap_appendix}

For a continuous chain distribution of $N$ ions with length $(N-1)d \approx Nd$, the effective potential required to maintain the uniform separation is \cite{leungEntanglingArbitraryPair2018},
\begin{equation}
	V_{\rm{trap}}(z) = \frac{q^{2}}{d}\ln \left(\frac{(Nd)^{2}}{(Nd)^{2} - (2z/d)^{2}}\right)~.
\end{equation}
Expanding this expression up to 4th-order around $z = 0$, 
\begin{equation}
	V_{\rm{trap}} \approx \frac{4 q^{2}}{d^{3}N^{2}}z^{2} + \frac{8q^{2}}{N^{4}d^{5}}z^{4}~,\label{eq: power_series_potential}
\end{equation}
the quadratic and quartic potential terms are then given by
\begin{subequations}
    \label{eq:kappa_params}
    \begin{align}
    \kappa_{2} &= \frac{8q^{2}}{d^{3}N^{2}}~,\label{eq:harmonic_term}\, \\
    \kappa_{4}&= \frac{32q^{2}}{d^{5}N^{4}}\,.
\end{align}
\end{subequations}

Due to the power series expansion, the inter-ion separation will no longer be uniform and the assumed value of $d$ underestimates the actual minimum inter-ion separations, which will increase for larger numbers of ions. Therefore, to ensure an equivalent comparison between ion chains of varying length, we reduce $d$ in Eq.~\eqref{eq:kappa_params} as the number of ions increases such that the minimum ion separation will be kept constant at \SI{3}{\micro\meter}.

\section{Monte-Carlo error model}\label{sec:Monte-Carlo_Errors}

To examine the performance of the fast gate mechanism against SDK pulse area errors, we employ a Monte-Carlo technique that stochastically samples a large ensemble of SDK sequences; in a given sequence, each kick has a probability $\epsilon$ of being erroneous. In order to enable the description of wavepacket diffraction into multiple momentum states, we assume that each element of the ideal kick sequence $\{z_k^0\}$ is perturbed according to the discrete stochastic process
\begin{align}
\label{eq:MCSampling_zk}
    z_k = \begin{cases}
        z_k^0 \, & \textrm{with probability } 1-\epsilon \\
        z_k^0 + n &  \textrm{with probability } \epsilon P_n
    \end{cases}
\end{align}
where the integers $n$ label the discrete probability distribution $P_n$ which satisfies $\sum_n P_n = 1$ and $P_0 = 0$. For each stochastically-sampled sequence $\{z_k\}$, the 2Q gate error can be computed from Eq.~\eqref{eq:infidelity}, which includes errors from both residual spin-motional entanglement \eqref{eq:entangling_phase} and imperfect phase accumulation \eqref{eq:motional_displacement}. We can then compute the probability distribution of the 2Q gate fidelity from a large ensemble of sampled errors, which can be done numerically. For the purposes of this work, we will consider a simplified error model of SDK errors that has two main error channels with equal probability: (1) a `no kick' corresponding to $n=-1$ with $P_{-1}=0.5$; and (2) a `backwards kick' corresponding to $n=-2$ with $P_{-2}=0.5$. These are two of the most significant error channels for Raman SDKs~\cite{mizrahiQuantumControlQubits2014}. 

For efficient numerical implementation, we avoid simulating trivial SDK sequences where there is no error in the pulse sequence. Instead, we simulate only the SDK sequences with at least one erroneous kick and include the trivial case of no errors when reconstructing the probability density function (PDF) of the gate error from the sampled 2Q gate errors. Furthermore, our numerical calculations use an error-class decomposition where the gate error PDF is calculated conditioned on there being exactly $m$ errors in the SDK sequence, i.e. $P_m(\varepsilon_{\rm G})$. The total gate error distribution can then be calculated by weighting the binomial likelihood of each error class:
\begin{align}
\label{eq:PDF_Binomial_Reconstruction}
    P(\varepsilon_{\rm G}) = \sum_{m=0}^{\mathcal{N}} \binom{\mathcal{N}}{m} \epsilon^m (1-\epsilon)^{\mathcal{N}-m} P_m(\varepsilon_{\rm G}) \,,
\end{align}
where the binomial coefficient $\binom{\mathcal{N}}{m}$ encodes the number of ways $m$ errors can be distributed amongst $\mathcal{N}$ kicks, and the factor $\epsilon^m (1-\epsilon)^{\mathcal{N}-m}$ encodes the probability of exactly $m$ errors occurring.

Our numerical calculations obey the following approach:
\begin{enumerate}
    \item First consider the distribution of 2Q gate errors conditioned on exactly $m=1$ SDK error in the sequence, $P_1(\varepsilon_{\rm G})$.
    \item Generate $S_1=10^4$ samples of errant SDK sequences, $\{ z_k\}$, following the rule Eq.~\eqref{eq:MCSampling_zk}.
    \item For each errant SDK sequence, compute the 2Q gate error from Eq.~\eqref{eq:infidelity}. This gives a set of $S_1$ samples of $\{\varepsilon_{\rm G}\}$ for the $m=1$ error order.
    \item Estimate the conditional PDF by sorting the samples of $\varepsilon_{\rm G}$ into $N_{\rm bins}$ with width $\Delta x$, i.e. 
    \begin{align}
        P_1(\varepsilon_{\rm G}) \approx (1/S_1)\sum_{{\rm bin}~i} \frac{{\rm count}^{(i)}}{\Delta x^{(i)}} \chi_i(\varepsilon_{\rm G})  \,,  \end{align} 
    where ${\rm count}^{(i)}$ is the number of data samples in the $i$-th bin with width $\Delta \varepsilon^{(i)}=\varepsilon^{(i+1)}-\varepsilon^{(i)}$ described by the indicator function: $\chi_i(\varepsilon)=1/\Delta \varepsilon^{(i)}$ for $\varepsilon\in [\varepsilon^{(i)},\varepsilon^{(i+1)}]$ and zero elsewhere.
    \item Repeat steps 1-4 for increasing error order $m\leq m_{\rm max}\leq \mathcal{N}$. We use $m_{\rm max}=3$, which we find to be sufficient for computing the full 2Q error PDF in the regime of small pulse error probability, i.e. $\epsilon \ll 1$.
    \item Reconstruct the full error PDF from Eq.~\eqref{eq:PDF_Binomial_Reconstruction}, truncating the sum at $m=m_{\rm max}$.
\end{enumerate}

\section{Derivation of fast gate unitary}\label{sec:unitary_derivation_appendix}
As operators on different motional modes commute, we will only consider the evolution of a single motional mode during the fast gate operation in this Appendix. For notational simplicity, we omit the explicit time-ordering operator.

From Eq.~\eqref{eq:GateUnitaryFull}, the unitary describing the evolution of a single motional mode during a 2Q fast gate is
\begin{align}
\label{eq:GateUnitaryFull_Appendix}
    \hat{\mathcal{U}}_{\rm G, \alpha} &=\prod_{m = A,B}\prod_{j=1}^{\mathcal{N}}\hat{\sigma}_{x}^{(m)}e^{i\kappa_{j}\phi^{(m)}(t_{j})\hat{\sigma}_{z}^{(m)}}\notag\\ & \qquad \qquad \times\hat{D}_{\alpha}\left(i\kappa_{j}\eta_{\alpha}b_{\alpha}^{(m)}\hat{\sigma}_{z}^{(m)}e^{i\omega_{\alpha} t_{j}} \right) \,,\\
    &=\prod_{m = A,B}\prod_{j=1}^{\mathcal{N}} \hat{\sigma}_{x}^{(m)}e^{i\hat{\gamma}_{j, \alpha}^{(m)}\hat{\sigma}_{z}^{(m)}}
\end{align}
where we have introduced
\begin{equation}\hat{\gamma}^{(m)}_{j, \alpha}(t_{j}) = \kappa_{j}\left[\phi(t_{j})^{(m)} + \eta_{\alpha}b_{\alpha}^{(m)}\left(e^{i\omega_{\alpha}t_{j}}\hat{a}_{\alpha}^{\dagger} + e^{-i\omega_{\alpha}t_{j}}\hat{a}_{\alpha}\right)\right]~.\end{equation}
We will explicitly consider the evolution described by the first two SDKs, 
\begin{align}
\label{eq:twoSDKs_Appendix}
    \hat{\mathcal{U}}_{\alpha} = \prod_{m = A,B}\hat{\sigma}_{x}^{(m)}e^{i\hat{\gamma}_{2, \alpha}^{(m)}(t_{2})\hat{\sigma}_{z}^{(m)}}\hat{\sigma}_{x}^{(m)}e^{i\hat{\gamma}_{1, \alpha}^{(m)}(t_{1})\hat{\sigma}_{z}^{(m)}} \,,
\end{align}
We can expand the second SDK, $e^{i\gamma\hat{\sigma}_{z}} =  \hat{I}\cos(\gamma) + i\hat{\sigma}_{z}\sin(\gamma)$ and by applying products of Pauli Matrices, we write this unitary as
\begin{align}
    \notag \hat{\mathcal{U}}_{\alpha}&=\prod_{m = A,B} \left(\hat{\sigma}_{x}^{(m)}\right)^{2}\left[\hat{I}\cos(\hat{\gamma}_{2, \alpha}(t_{2})) \right.\\&\qquad \qquad \left.-i\hat{\sigma}^{(m)}_{z}\sin(\hat{\gamma}_{2,\alpha}(t_{2}))\right] e^{i\hat{\gamma}_{1, \alpha}^{(m)}(t_{1})\hat{\sigma}_{z}^{(m)}}~,\\
    &=\prod_{m = A,B} \left(\hat{\sigma}_{x}^{(m)}\right)^{2}e^{-i\hat{\gamma}_{2,\alpha}^{(m)}\hat{\sigma}_{z}^{(m)}}e^{i\hat{\gamma}_{1,\alpha}^{(m)}\hat{\sigma}_{z}^{(m)}}~,\\
    &= \prod_{m = A,B} \prod_{j = 1}^{2}\left(\hat{\sigma}_{x}^{(m)}\right)^{2}e^{i(-1)^{j+1}\kappa_{j} \phi^{(m)}(t_{j})\hat{\sigma}_{z}^{(m)}}\notag\\&\qquad\qquad\times\hat{D}_{\alpha}\left(i(-1)^{j+1}\kappa_{j}\eta_{\alpha}b_{\alpha}^{(m)}\hat{\sigma}_{z}^{(m)}e^{i\omega_{\alpha}t_{j}}\right)\label{eq:final_twoSDKs_Appendix}
\end{align}
We will now consider the product of time-ordered displacement operators. Using the property of the product of displacement operators $\hat{D}(a)\hat{D}(b) = e^{i\rm{Im}[ab^{*}]}\hat{D}(a+b)$, we find:
\begin{align}
    \prod_{j = 1}^{2}\hat{D}_{\alpha}\left(i\eta_{\alpha, j}^{(AB)}\right)=e^{i\rm{Im\left[\eta_{\alpha,2}^{(AB)}\left(\eta_{\alpha, 1}^{(AB)}\right)^{*}\right]}} \hat{D}_{\alpha}\left[i\left(\sum_{j = 1}^{2}\eta_{\alpha, j}^{(AB)}\right)\right] \label{eq:two_displacements}
\end{align}
where $\eta^{(AB)}_{\alpha,j} = (-1)^{j+1}\kappa_{j}\eta_{\alpha} (b^{(A)}_{\alpha}\hat{\sigma}_{z}^{(A)} + b^{(B)}_{\alpha}\hat{\sigma}_{z}^{(B)})e^{i\omega_{\alpha}t_{j}}$.

Repeating Eq.~\eqref{eq:twoSDKs_Appendix} - ~\eqref{eq:two_displacements}, the single-mode gate unitary takes the form,
\begin{align}
    \hat{\mathcal{U}}_{\rm G, \alpha} &=\left(\hat{\sigma}_{x}^{(A)}\hat{\sigma}_{x}^{(B)}\right)^{\mathcal{N}}\prod_{j=1}^{\mathcal{N}}e^{i(-1)^{j+1}\kappa_{j}\left(\phi^{(A)}(t_{j})\hat{\sigma}_{z}^{(A)} + \phi^{(B)}(t_{j})\hat{\sigma}_{z}^{(B)}\right)} \notag \\& \times e^{i\sum_{j = 1}^{\mathcal{N}}\rm{Im\left[\eta_{\alpha,j}^{(AB)}\left(\sum_{k = 1}^{j-1}\eta_{\alpha, k}^{(AB)}\right)^{*}\right]}} \hat{D}_{\alpha}\left[i\left(\sum_{j = 1}^{\mathcal{N}}\eta_{\alpha, j}^{(AB)}\right)\right]\,.
\end{align}

\section{ODE description of gate dynamics}\label{sec:ode_appendix}

Fast gate dynamics can be described using an ODE description of the classical motion of the trapped ions for each state-dependent trajectory. The motion in a chain of ions is governed by the trapping potential and the Coulomb interaction between ions, $V(z) = \sum_{j = 1}^{N}(\kappa_{2}z_{j}^{2}/2 + \kappa_{4}z_{j}^{2}/4) +\sum_{j \neq k}^{N}q^{2}/|z_{j} - z_{k}|$. The Coulomb interaction couples the ion motion, which can be effectively described by decomposing the collective dynamics into the motional modes of the system. The exact mode structure is expensive to compute due to the complex non-linear dynamics in multi-ion chains. Therefore, assuming each ion only experiences small-displacements from equilibirum, $z_{0}$, we describe the ion motion as harmonic oscillations of normal modes by linearising the Coulomb interaction \cite{jamesQuantumDynamicsCold1998}. The coupling vector, $\bm{b}_{\alpha}$ and oscillation frequency, $\omega_{\alpha}$ of each mode can be calculated from the eigenvalue equation,
\begin{equation}
    \bm{H} \cdot \bm{b}_{\alpha} = \omega_{\alpha}^{2}\bm{b}_{\alpha}~,
\end{equation}
where $\bm{H}$ is the Hessian matrix with elements $H_{jk} = \frac{1}{M} \frac{\partial V}{\partial z_{j}\partial z_{k}}|_{z_{0}}$. 

We quantise the dynamics by treating each mode as an independent quantum harmonic oscillator described by the Hamiltonian,
\begin{equation}
    \hat{H}_{\alpha} = \hbar \omega_{\alpha}\left(\hat{a}^{\dagger}_{\alpha}\hat{a}_{\alpha}+\frac{1}{2}\right)~,
\end{equation}
where $\hat{a}_{\alpha}$ ($\hat{a}^{\dagger}_{\alpha}$) is the annihilation (creation) operator described in terms of the dimensionless position and momentum quadratures, $\hat{X}_{\alpha}$ and $\hat{Y}_{\alpha}$,
\begin{subequations}
    \label{eq:ModeQuadratures}
    \begin{align}
    \hat{a}_{\alpha} &= \frac{1}{\sqrt{2}}(\hat{X}_{\alpha} + i\hat{Y}_{\alpha})\,, \\
    \hat{a}^{\dagger}_{\alpha} &= \frac{1}{\sqrt{2}}(\hat{X}_{\alpha} - i\hat{Y}_{\alpha})\,,
\end{align}
\end{subequations}
Computing the dynamics of the system in the Heisenberg picture $i\hbar \frac{d \hat{X}_{\alpha}}{dt} = [\hat{X}_{\alpha}, \hat{H}]$ yields
\begin{subequations}
\label{eq:HeisenbergEOM_Quadratures}
    \begin{align}
\label{eq:EOM_Xhat}
     \frac{d\hat{X}_{\alpha}}{dt} &= \omega_{\alpha}\hat{Y}_{\alpha} \,, \\
     \label{eq:EOM_Yhat}
     \frac{d\hat{Y}_{\alpha}}{dt} &= -\omega_{\alpha}\hat{X}_{\alpha} \,.
\end{align}
\end{subequations}

Assuming each mode can be described by a coherent state, $\ket{\beta_{\alpha}}$, we have $\langle \beta_{\alpha}(t) | \hat{a}_{\alpha} | \beta_{\alpha}(t) \rangle = \beta_{\alpha} = \frac{1}{\sqrt{2}
}\left( \langle\hat{X}_{\alpha}(t)\rangle + i \langle\hat{Y}_{\alpha}(t) \rangle \right)$. Therefore, taking the expectation values of Eqs.~\eqref{eq:HeisenbergEOM_Quadratures} allows us to define the equations of motion of each normal mode,
\begin{subequations}
\label{eq:EOM_Quadratures}
    \begin{align}
\label{eq:EOM_X}
     \ddot{X}_{\alpha} &= -\omega^{2}_{\alpha}X_{\alpha}\,, \\
     \label{eq:EOM_Y}
     Y_{\alpha} &= \frac{\dot{X}_{\alpha}}{\omega_{\alpha}} \,,
\end{align}
\end{subequations}
where $X_{\alpha} = \langle\hat{X}_{\alpha}(t)\rangle$ and $Y_{\alpha} = \langle\hat{Y}_{\alpha}(t) \rangle$.
Eqs.~\eqref{eq:EOM_Quadratures} describe the free evolution of each mode between SDKs. 

For a gate between a pair of ions ($m$, $n$), we model each SDK as an instantaneous transform of the momentum quadrature, 
\begin{equation}
\label{eq:momentum_transform}
    Y_{\alpha} \rightarrow Y_{\alpha} \pm \sqrt{2} z_{j} b_{\alpha}^{(\pm)}\eta_{\alpha}~, 
\end{equation}
where the sign of the transformation depends on the two-qubit state at the start of the gate and
\begin{subequations}
    \begin{align}
     b_{\alpha}^{(+)} &= b_{\alpha}^{(m)} + b_{\alpha}^{(n)}\,, \\
     b_{\alpha}^{(-)} &= b_{\alpha}^{(m)} - b_{\alpha}^{(n)} \,,
\end{align}
\end{subequations}
are the effective coupling of the SDK to the same-spin ($\{\ket{\downarrow \downarrow}, \ket{\uparrow \uparrow}\}$) and opposite-spin $\{\ket{\downarrow \uparrow}, \ket{\uparrow \downarrow}\}$ two-qubit states respectively. 

Therefore, to fully describe the gate dynamics, Eq.~\eqref{eq:EOM_Quadratures} must be piecewise integrated with new initial conditions defined after each SDK for all initial two-qubit states $\{\ket{\downarrow \downarrow}, \ket{\downarrow \uparrow}, \ket{\uparrow \downarrow}, \ket{\uparrow \uparrow}\}$. Note that it is only necessary to evaluate the motional trajectories for $\{\ket{\downarrow \downarrow}, \ket{\downarrow \uparrow}\}$ as the trajectories for $\{\ket{\uparrow \downarrow}, \ket{\uparrow \uparrow}\}$ are given by symmetry \cite{mehdiFastMixedspeciesQuantum2025} ($X \rightarrow -X$ and $Y \rightarrow -Y$).

We calculate the phase accumulated at time $t_{j}$ due to a single SDK at time $t_{0}$ by considering the difference between the unkicked coherent state, $\ket{\beta_{\alpha}(t_{j})}$, 
\begin{equation}
    \beta_{\alpha}(t_{j}) = \frac{1}{\sqrt{2}}\left(X_{\alpha}(t_{0}) + iY_{\alpha}(t_{0})\right)e^{-i\omega_{\alpha}t_{j}}~,
\end{equation}
and the kicked coherent state, $\ket{\beta_{\alpha}'(t_{j})}$
\begin{equation}
    \beta_{\alpha}'(t_{j}) = \frac{1}{\sqrt{2}}\left(X_{\alpha}(t_{0}) + i(Y_{\alpha}(t_{0}) + \Delta Y_{\alpha})\right)e^{-i\omega_{\alpha}t_{j}}~,
\end{equation}
where $\Delta Y_{\alpha} = \pm \sqrt{2} z_{j}b_{\alpha}^{(\pm)}\eta_{\alpha}$ depends on the initial spin-state. Calculating the overlap of these states, $\langle\beta_{\alpha}(t_{j})|\beta_{\alpha}'(t_{j})\rangle$, we find 
\begin{equation}
    \theta_{\alpha}(t_{j+1}) = \frac{1}{2}X(t_{j})\Delta Y_{\alpha}~.
\end{equation}
To calculate the total accumulated phase from each mode for each two-qubit state, we perform a piecewise summation over each of the SDKs in the gate for each initial 2Q state: 
\begin{subequations}
    \begin{align}
    \theta_{\alpha, \ket{\uparrow\uparrow}} = \theta_{\alpha, \ket{\downarrow\downarrow}} &= \frac{1}{2}\sum_{j = 1}^{\mathcal{N}}\sqrt{2} z_{j}b_{\alpha}^{+}\eta_{\alpha}X_{\alpha, \ket{\uparrow\uparrow}}(t_{j})\,, \\
    \theta_{\alpha, \ket{\uparrow\downarrow}} = \theta_{\alpha, \ket{\downarrow\downarrow}} &= \frac{1}{2}\sum_{j = 1}^{\mathcal{N}}\sqrt{2} z_{j}b_{\alpha}^{-}\eta_{\alpha}X_{\alpha, \ket{\uparrow\downarrow}}(t_{j})\,. 
\end{align}
\end{subequations}

The gate error is given by 
\begin{align}
    \varepsilon_{\rm{av}} = \frac{2}{3}& \left\lvert\frac{\pi}{4} - \Theta\right\rvert^{2}\notag\\& + \frac{2}{3} \sum_{i,j}^{\{\uparrow, \downarrow\}}\sum_{\alpha=1}^{N} \left(\frac{1}{2} + \bar{n}_{\alpha}\right)\left\lvert\Delta \beta_{\alpha, \ket{i j }}\right\rvert^{2}~,\label{eq:ode_infidelity}
\end{align}
where
\begin{align}
    \Theta &= \frac{1}{4}\sum_{\alpha = 1}^{N}\left(\theta_{\alpha, \ket{\downarrow\downarrow}} + \theta_{\alpha, \ket{\uparrow\uparrow}} - \theta_{\alpha, \ket{\downarrow\uparrow}}- \theta_{\alpha, \ket{\uparrow\downarrow}}\right)
\end{align}
and
\begin{align}
    \left\lvert\Delta \beta_{\alpha,\ket{ij}}\right\rvert^{2} &= \frac{1}{2} \left( \Delta X_{\alpha,\ket{ij}}^{2} + \Delta Y_{\alpha,\ket{ij}}^{2}\right) 
\end{align}
are the final phase and motional displacement of each two-qubit state, respectively.

\bibliographystyle{bibsty}
\bibliography{Trapped_ions_bib}

@article{Gould1986,
  title = {Diffraction of atoms by light: The near-resonant Kapitza-Dirac effect},
  volume = {56},
  ISSN = {0031-9007},
  url = {http://dx.doi.org/10.1103/PhysRevLett.56.827},
  DOI = {10.1103/physrevlett.56.827},
  number = {8},
  journal = {Physical Review Letters},
  publisher = {American Physical Society (APS)},
  author = {Gould,  Phillip L. and Ruff,  George A. and Pritchard,  David E.},
  year = {1986},
  month = feb,
  pages = {827–830}
}

@misc{anProgrammableAdiabaticRapid2025,
  title = {Programmable {{Adiabatic Rapid Passage}} Laser Pulses for {{Ultra-fast Gates}} on Trapped Ions},
  author = {An, En-Teng and Zhang, Hao-Qing and Huang, Yun-Feng and Li, Chuan-Feng and Cui, Jin-Ming},
  year = 2025,
  month = nov,
  number = {arXiv:2511.04893},
  eprint = {2511.04893},
  primaryclass = {quant-ph},
  publisher = {arXiv},
  doi = {10.48550/arXiv.2511.04893},
  urldate = {2025-11-10},
  archiveprefix = {arXiv}
}

@article{pogorelovCompactIonTrapQuantum2021,
  title = {Compact {{Ion-Trap Quantum Computing Demonstrator}}},
  author = {Pogorelov, I. and Feldker, T. and Marciniak, {\relax Ch}. D. and Postler, L. and Jacob, G. and Krieglsteiner, O. and Podlesnic, V. and Meth, M. and Negnevitsky, V. and Stadler, M. and H{\"o}fer, B. and W{\"a}chter, C. and Lakhmanskiy, K. and Blatt, R. and Schindler, P. and Monz, T.},
  year = 2021,
  month = jun,
  journal = {PRX Quantum},
  volume = {2},
  number = {2},
  pages = {020343},
  issn = {2691-3399},
  doi = {10.1103/PRXQuantum.2.020343},
  urldate = {2025-11-17},
  langid = {english}
}

@misc{liuHighFidelityRamanSpinDependent2025,
  title = {High-{{Fidelity Raman Spin-Dependent Kicks}} in the {{Presence}} of {{Micromotion}}},
  author = {Liu, Haonan and Vaidya, Varun D. and Galan, Monica Gutierrez and Ratcliffe, Alexander K. and Poudel, Amrit and Viteri, C. Ricardo},
  year = 2025,
  month = nov,
  number = {arXiv:2511.15959},
  eprint = {2511.15959},
  primaryclass = {quant-ph},
  publisher = {arXiv},
  doi = {10.48550/arXiv.2511.15959},
  urldate = {2025-12-06},
  archiveprefix = {arXiv}
}

@article{fengEfficientGroundStateCooling2020,
  title = {Efficient {{Ground-State Cooling}} of {{Large Trapped-Ion Chains}} with an {{Electromagnetically-Induced-Transparency Tripod Scheme}}},
  author = {Feng, L. and Tan, W. L. and De, A. and Menon, A. and Chu, A. and Pagano, G. and Monroe, C.},
  year = {2020},
  month = jul,
  journal = {Physical Review Letters},
  volume = {125},
  number = {5},
  pages = {053001},
  issn = {0031-9007, 1079-7114},
  doi = {10.1103/PhysRevLett.125.053001},
  urldate = {2025-08-06},
  langid = {english}
}

@dataset{savill-brownDatasetErrorResilientFast2025,
  title = {Dataset for "{{Error-Resilient Fast Entangling Gates}} for {{Scalable Ion-Trap Quantum Processors}}"},
  author = {Savill-Brown, Isabelle},
  date = {2025-08-12},
  publisher = {Zenodo},
  url = {https://zenodo.org/records/16809845},
  urldate = {2025-08-12}
}

@article{niedermayrCryogenicSurfaceIon2014,
  title = {Cryogenic Surface Ion Trap Based on Intrinsic Silicon},
  author = {Niedermayr, Michael and Lakhmanskiy, Kirill and Kumph, Muir and Partel, Stefan and Edlinger, Jonannes and Brownnutt, Michael and Blatt, Rainer},
  year = {2014},
  month = nov,
  journal = {New Journal of Physics},
  volume = {16},
  number = {11},
  pages = {113068},
  publisher = {IOP Publishing},
  issn = {1367-2630},
  doi = {10.1088/1367-2630/16/11/113068},
  urldate = {2025-08-08},
  langid = {english}
}

@article{wuElectromagneticallyinducedtransparencyCoolingTripod2025,
  title = {Electromagnetically-Induced-Transparency Cooling with a Tripod Structure in a Hyperfine Trapped Ion with Mixed-Species Crystals},
  author = {Wu, Jenny J. and Hou, Pan-Yu and Erickson, Stephen D. and Brandt, Adam D. and Wan, Yong and Zarantonello, Giorgio and Cole, Daniel C. and Wilson, Andrew C. and Slichter, Daniel H. and Leibfried, Dietrich},
  year = {2025},
  month = apr,
  journal = {Physical Review A},
  volume = {111},
  number = {4},
  pages = {043109},
  issn = {2469-9926, 2469-9934},
  doi = {10.1103/PhysRevA.111.043109},
  urldate = {2025-08-06},
  langid = {english}
}

@online{QuantinuumDominatesQuantum,
  title = {Quantinuum {{Dominates}} the {{Quantum Landscape}}: {{New World-Record}} in {{Quantum Volume}}},
  shorttitle = {Quantinuum {{Dominates}} the {{Quantum Landscape}}},
  url = {https://www.quantinuum.com/blog/quantum-volume-milestone},
  urldate = {2025-07-31},
  langid = {english}
}

@article{smithSingleQubitGatesErrors2025,
  title = {Single-{{Qubit Gates}} with {{Errors}} at the \$\{10\}{\textasciicircum}\{{\textbackslash}ensuremath\{-\}7\}\$ {{Level}}},
  author = {Smith, M. C. and Leu, A. D. and Miyanishi, K. and Gely, M. F. and Lucas, D. M.},
  year = {2025},
  month = jun,
  journal = {Physical Review Letters},
  volume = {134},
  number = {23},
  pages = {230601},
  publisher = {American Physical Society},
  doi = {10.1103/42w2-6ccy},
  urldate = {2025-07-23}
}

@inproceedings{muraliArchitectingNoisyIntermediateScale2020,
  title = {Architecting {{Noisy Intermediate-Scale Trapped Ion Quantum Computers}}},
  booktitle = {2020 {{ACM}}/{{IEEE}} 47th {{Annual International Symposium}} on {{Computer Architecture}} ({{ISCA}})},
  author = {Murali, Prakash and Debroy, Dripto M. and Brown, Kenneth R. and Martonosi, Margaret},
  year = {2020},
  month = may,
  pages = {529--542},
  doi = {10.1109/ISCA45697.2020.00051},
  urldate = {2025-07-13}
}

@article{nigmatullinMinimallyComplexIon2016,
  title = {Minimally Complex Ion Traps as Modules for Quantum Communication and Computing},
  author = {Nigmatullin, Ramil and Ballance, Christopher J and de Beaudrap, Niel and Benjamin, Simon C},
  year = {2016},
  month = oct,
  journal = {New Journal of Physics},
  volume = {18},
  number = {10},
  pages = {103028},
  publisher = {IOP Publishing},
  issn = {1367-2630},
  doi = {10.1088/1367-2630/18/10/103028},
  urldate = {2025-07-13},
  langid = {english}
}

@article{nickersonFreelyScalableQuantum2014,
  title = {Freely {{Scalable Quantum Technologies Using Cells}} of 5-to-50 {{Qubits}} with {{Very Lossy}} and {{Noisy Photonic Links}}},
  author = {Nickerson, Naomi H. and Fitzsimons, Joseph F. and Benjamin, Simon C.},
  year = {2014},
  month = dec,
  journal = {Physical Review X},
  volume = {4},
  number = {4},
  pages = {041041},
  publisher = {American Physical Society},
  doi = {10.1103/PhysRevX.4.041041},
  urldate = {2025-07-13}
}

@article{schoenbergerShuttlingScalableTrappedIon2025,
  title = {Shuttling for {{Scalable Trapped-Ion Quantum Computers}}},
  author = {Schoenberger, Daniel and Hillmich, Stefan and Brandl, Matthias and Wille, Robert},
  year = {2025},
  month = jun,
  journal = {IEEE Transactions on Computer-Aided Design of Integrated Circuits and Systems},
  volume = {44},
  number = {6},
  pages = {2144--2155},
  issn = {1937-4151},
  doi = {10.1109/TCAD.2024.3513262},
  urldate = {2025-07-13}
}

@article{sterkClosedloopOptimizationFast2022,
  title = {Closed-Loop Optimization of Fast Trapped-Ion Shuttling with Sub-Quanta Excitation},
  author = {Sterk, Jonathan D. and Coakley, Henry and Goldberg, Joshua and Hietala, Vincent and Lechtenberg, Jason and McGuinness, Hayden and McMurtrey, Daniel and Parazzoli, L. Paul and Van Der Wall, Jay and Stick, Daniel},
  year = {2022},
  month = jun,
  journal = {npj Quantum Information},
  volume = {8},
  number = {1},
  pages = {68},
  publisher = {Nature Publishing Group},
  issn = {2056-6387},
  doi = {10.1038/s41534-022-00579-3},
  urldate = {2025-07-13},
  copyright = {2022 The Author(s)},
  langid = {english}
}

@article{mordiniMultizoneTrappedIonQubit2025,
  title = {Multizone {{Trapped-Ion Qubit Control}} in an {{Integrated Photonics QCCD Device}}},
  author = {Mordini, Carmelo and Ricci Vasquez, Alfredo and Motohashi, Yuto and M{\"u}ller, Mose and Malinowski, Maciej and Zhang, Chi and Mehta, Karan K. and Kienzler, Daniel and Home, Jonathan P.},
  year = {2025},
  month = feb,
  journal = {Physical Review X},
  volume = {15},
  number = {1},
  pages = {011040},
  publisher = {American Physical Society},
  doi = {10.1103/PhysRevX.15.011040},
  urldate = {2025-07-13}
}

@misc{sahaHighfidelityRemoteEntanglement2024,
  title = {High-Fidelity Remote Entanglement of Trapped Atoms Mediated by Time-Bin Photons},
  author = {Saha, Sagnik and Shalaev, Mikhail and O'Reilly, Jameson and Goetting, Isabella and Toh, George and Kalakuntla, Ashish and Yu, Yichao and Monroe, Christopher},
  year = {2024},
  month = jun,
  number = {arXiv:2406.01761},
  eprint = {2406.01761},
  primaryclass = {quant-ph},
  publisher = {arXiv},
  doi = {10.48550/arXiv.2406.01761},
  urldate = {2025-07-13},
  archiveprefix = {arXiv}
}

@article{oreillyFastPhotonMediatedEntanglement2024,
  title = {Fast {{Photon-Mediated Entanglement}} of {{Continuously Cooled Trapped Ions}} for {{Quantum Networking}}},
  author = {O'Reilly, Jameson and Toh, George and Goetting, Isabella and Saha, Sagnik and Shalaev, Mikhail and Carter, Allison L. and Risinger, Andrew and Kalakuntla, Ashish and Li, Tingguang and Verma, Ashrit and Monroe, Christopher},
  year = {2024},
  month = aug,
  journal = {Physical Review Letters},
  volume = {133},
  number = {9},
  pages = {090802},
  publisher = {American Physical Society},
  doi = {10.1103/PhysRevLett.133.090802},
  urldate = {2025-07-13}
}

@article{drmotaRobustQuantumMemory2023,
  title = {Robust {{Quantum Memory}} in a {{Trapped-Ion Quantum Network Node}}},
  author = {Drmota, P. and Main, D. and Nadlinger, D. P. and Nichol, B. C. and Weber, M. A. and Ainley, E. M. and Agrawal, A. and Srinivas, R. and Araneda, G. and Ballance, C. J. and Lucas, D. M.},
  year = {2023},
  month = mar,
  journal = {Physical Review Letters},
  volume = {130},
  number = {9},
  pages = {090803},
  publisher = {American Physical Society},
  doi = {10.1103/PhysRevLett.130.090803},
  urldate = {2025-07-13}
}

@article{palmeroFastPhaseGates2017,
  title = {Fast Phase Gates with Trapped Ions},
  author = {Palmero, M. and {Mart{\'i}nez-Garaot}, S. and Leibfried, D. and Wineland, D. J. and Muga, J. G.},
  year = {2017},
  month = feb,
  journal = {Physical Review A},
  volume = {95},
  number = {2},
  pages = {022328},
  publisher = {American Physical Society},
  doi = {10.1103/PhysRevA.95.022328},
  urldate = {2025-07-13}
}

@article{johnsonSensingAtomicMotion2015,
  title = {Sensing {{Atomic Motion}} from the {{Zero Point}} to {{Room Temperature}} with {{Ultrafast Atom Interferometry}}},
  author = {Johnson, K. G. and Neyenhuis, B. and Mizrahi, J. and {Wong-Campos}, J. D. and Monroe, C.},
  year = {2015},
  month = nov,
  journal = {Physical Review Letters},
  volume = {115},
  number = {21},
  pages = {213001},
  publisher = {American Physical Society},
  doi = {10.1103/PhysRevLett.115.213001},
  urldate = {2025-07-11}
}

@article{cumminsTacklingSystematicErrors2003,
  title = {Tackling Systematic Errors in Quantum Logic Gates with Composite Rotations},
  author = {Cummins, Holly K. and Llewellyn, Gavin and Jones, Jonathan A.},
  year = {2003},
  month = apr,
  journal = {Physical Review A},
  volume = {67},
  number = {4},
  pages = {042308},
  publisher = {American Physical Society},
  doi = {10.1103/PhysRevA.67.042308},
  urldate = {2025-07-11}
}

@article{bergmannRoadmapSTIRAPApplications2019,
  title = {Roadmap on {{STIRAP}} Applications},
  author = {Bergmann, Klaas and N{\"a}gerl, Hanns-Christoph and Panda, Cristian and Gabrielse, Gerald and Miloglyadov, Eduard and Quack, Martin and Seyfang, Georg and Wichmann, Gunther and Ospelkaus, Silke and Kuhn, Axel and Longhi, Stefano and Szameit, Alexander and Pirro, Philipp and Hillebrands, Burkard and Zhu, Xue-Feng and Zhu, Jie and Drewsen, Michael and Hensinger, Winfried K and Weidt, Sebastian and Halfmann, Thomas and Wang, Hai-Lin and Paraoanu, Gheorghe Sorin and Vitanov, Nikolay V and Mompart, Jordi and Busch, Thomas and Barnum, Timothy J and Grimes, David D and Field, Robert W and Raizen, Mark G and Narevicius, Edvardas and Auzinsh, Marcis and Budker, Dmitry and P{\'a}lffy, Adriana and Keitel, Christoph H},
  year = {2019},
  month = sep,
  journal = {Journal of Physics B: Atomic, Molecular and Optical Physics},
  volume = {52},
  number = {20},
  pages = {202001},
  publisher = {IOP Publishing},
  issn = {0953-4075},
  doi = {10.1088/1361-6455/ab3995},
  urldate = {2025-07-11},
  langid = {english}
}

@article{akhtarHighfidelityQuantumMatterlink2023,
  title = {A High-Fidelity Quantum Matter-Link between Ion-Trap Microchip Modules},
  author = {Akhtar, M. and Bonus, F. and {Lebrun-Gallagher}, F. R. and Johnson, N. I. and {Siegele-Brown}, M. and Hong, S. and Hile, S. J. and Kulmiya, S. A. and Weidt, S. and Hensinger, W. K.},
  year = {2023},
  month = feb,
  journal = {Nature Communications},
  volume = {14},
  number = {1},
  pages = {531},
  publisher = {Nature Publishing Group},
  issn = {2041-1723},
  doi = {10.1038/s41467-022-35285-3},
  urldate = {2025-06-30},
  copyright = {2023 The Author(s)},
  langid = {english}
}

@article{anHighFidelityState2022a,
  title = {High Fidelity State Preparation and Measurement of Ion Hyperfine Qubits with {{I}} {$>$} 1/2},
  author = {An, Fangzhao Alex and Ransford, Anthony and Schaffer, Andrew and Sletten, Lucas R. and Gaebler, John and Hostetter, James and Vittorini, Grahame},
  year = {2022},
  month = sep,
  journal = {Physical Review Letters},
  volume = {129},
  number = {13},
  eprint = {2203.01920},
  primaryclass = {quant-ph},
  pages = {130501},
  issn = {0031-9007, 1079-7114},
  doi = {10.1103/PhysRevLett.129.130501},
  urldate = {2024-04-06},
  archiveprefix = {arXiv}
}

@article{bentleyFastGatesIon2013,
  title = {Fast Gates for Ion Traps by Splitting Laser Pulses},
  author = {Bentley, C. D. B. and Carvalho, A. R. R. and Kielpinski, D. and Hope, J. J.},
  year = {2013},
  month = apr,
  journal = {New Journal of Physics},
  volume = {15},
  number = {4},
  eprint = {1211.7156},
  primaryclass = {quant-ph},
  pages = {043006},
  issn = {1367-2630},
  doi = {10.1088/1367-2630/15/4/043006},
  urldate = {2024-04-12},
  archiveprefix = {arXiv}
}

@article{bentleyStabilityThresholdsCalculation2016,
  title = {Stability Thresholds and Calculation Techniques for Fast Entangling Gates on Trapped Ions},
  author = {Bentley, C. D. B. and Taylor, R. L. and Carvalho, A. R. R. and Hope, J. J.},
  year = {2016},
  month = apr,
  journal = {Physical Review A},
  volume = {93},
  number = {4},
  pages = {042342},
  issn = {2469-9926, 2469-9934},
  doi = {10.1103/PhysRevA.93.042342},
  urldate = {2024-07-18},
  copyright = {http://link.aps.org/licenses/aps-default-license},
  langid = {english}
}

@article{bentleyTrappedIonScaling2015,
  title = {Trapped Ion Scaling with Pulsed Fast Gates},
  author = {Bentley, C. D. B. and Carvalho, A. R. R. and Hope, J. J.},
  year = {2015},
  month = oct,
  journal = {New Journal of Physics},
  volume = {17},
  number = {10},
  pages = {103025},
  publisher = {IOP Publishing},
  issn = {1367-2630},
  doi = {10.1088/1367-2630/17/10/103025},
  urldate = {2024-07-19},
  langid = {english}
}

@article{bruzewiczMeasurementIonMotional2015,
  title = {Measurement of Ion Motional Heating Rates over a Range of Trap Frequencies and Temperatures},
  author = {Bruzewicz, C. D. and Sage, J. M. and Chiaverini, J.},
  year = {2015},
  month = apr,
  journal = {Physical Review A},
  volume = {91},
  number = {4},
  pages = {041402},
  issn = {1050-2947, 1094-1622},
  doi = {10.1103/PhysRevA.91.041402},
  urldate = {2024-10-11},
  copyright = {http://link.aps.org/licenses/aps-default-license},
  langid = {english}
}

@article{bruzewiczTrappedIonQuantumComputing2019,
  title = {Trapped-{{Ion Quantum Computing}}: {{Progress}} and {{Challenges}}},
  shorttitle = {Trapped-{{Ion Quantum Computing}}},
  author = {Bruzewicz, Colin D. and Chiaverini, John and McConnell, Robert and Sage, Jeremy M.},
  year = {2019},
  month = jun,
  journal = {Applied Physics Reviews},
  volume = {6},
  number = {2},
  eprint = {1904.04178},
  primaryclass = {physics, physics:quant-ph},
  pages = {021314},
  issn = {1931-9401},
  doi = {10.1063/1.5088164},
  urldate = {2024-03-31},
  archiveprefix = {arXiv}
}

@article{campbellUltrafastGatesSingle2010,
  title = {Ultrafast {{Gates}} for {{Single Atomic Qubits}}},
  author = {Campbell, W. C. and Mizrahi, J. and Quraishi, Q. and Senko, C. and Hayes, D. and Hucul, D. and Matsukevich, D. N. and Maunz, P. and Monroe, C.},
  year = {2010},
  month = aug,
  journal = {Physical Review Letters},
  volume = {105},
  number = {9},
  pages = {090502},
  issn = {0031-9007, 1079-7114},
  doi = {10.1103/PhysRevLett.105.090502},
  urldate = {2024-08-12},
  copyright = {http://link.aps.org/licenses/aps-default-license},
  langid = {english}
}

@article{cetinaControlTransverseMotion2022,
  title = {Control of {{Transverse Motion}} for {{Quantum Gates}} on {{Individually Addressed Atomic Qubits}}},
  author = {Cetina, M. and Egan, L.N. and Noel, C. and Goldman, M.L. and Biswas, D. and Risinger, A.R. and Zhu, D. and Monroe, C.},
  year = {2022},
  month = mar,
  journal = {PRX Quantum},
  volume = {3},
  number = {1},
  pages = {010334},
  issn = {2691-3399},
  doi = {10.1103/PRXQuantum.3.010334},
  urldate = {2024-08-05},
  langid = {english}
}

@article{charlesdoretControllingTrappingPotentials2012,
  title = {Controlling Trapping Potentials and Stray Electric Fields in a Microfabricated Ion Trap through Design and Compensation},
  author = {Charles Doret, S and Amini, Jason M and Wright, Kenneth and Volin, Curtis and Killian, Tyler and Ozakin, Arkadas and Denison, Douglas and Hayden, Harley and Pai, C-S and Slusher, Richart E and Harter, Alexa W},
  year = {2012},
  month = jul,
  journal = {New Journal of Physics},
  volume = {14},
  number = {7},
  pages = {073012},
  publisher = {IOP Publishing},
  issn = {1367-2630},
  doi = {10.1088/1367-2630/14/7/073012},
  urldate = {2025-06-04},
  langid = {english}
}

@misc{chenBenchmarkingTrappedionQuantum2023,
  title = {Benchmarking a Trapped-Ion Quantum Computer with 29 Algorithmic Qubits},
  author = {Chen, Jwo-Sy and Nielsen, Erik and Ebert, Matthew and Inlek, Volkan and Wright, Kenneth and Chaplin, Vandiver and Maksymov, Andrii and P{\'a}ez, Eduardo and Poudel, Amrit and Maunz, Peter and Gamble, John},
  year = {2023},
  month = aug,
  number = {arXiv:2308.05071},
  eprint = {2308.05071},
  publisher = {arXiv},
  urldate = {2024-10-15},
  archiveprefix = {arXiv}
}

@article{choiOptimalQuantumControl2014,
  title = {Optimal {{Quantum Control}} of {{Multimode Couplings}} between {{Trapped Ion Qubits}} for {{Scalable Entanglement}}},
  author = {Choi, T. and Debnath, S. and Manning, T. A. and Figgatt, C. and Gong, Z.-X. and Duan, L.-M. and Monroe, C.},
  year = {2014},
  month = may,
  journal = {Physical Review Letters},
  volume = {112},
  number = {19},
  pages = {190502},
  publisher = {American Physical Society},
  doi = {10.1103/PhysRevLett.112.190502},
  urldate = {2025-06-10}
}

@article{ciracQuantumComputationsCold1995,
  title = {Quantum {{Computations}} with {{Cold Trapped Ions}}},
  author = {Cirac, J. I. and Zoller, P.},
  year = {1995},
  month = may,
  journal = {Physical Review Letters},
  volume = {74},
  number = {20},
  pages = {4091--4094},
  issn = {0031-9007, 1079-7114},
  doi = {10.1103/PhysRevLett.74.4091},
  urldate = {2024-08-12},
  langid = {english}
}

@article{clarkHighFidelityBellStatePreparation2021,
  title = {High-{{Fidelity Bell-State Preparation}} with {{Ca40}}+ {{Optical Qubits}}},
  author = {Clark, Craig R. and Tinkey, Holly N. and Sawyer, Brian C. and Meier, Adam M. and Burkhardt, Karl A. and Seck, Christopher M. and Shappert, Christopher M. and Guise, Nicholas D. and Volin, Curtis E. and Fallek, Spencer D. and Hayden, Harley T. and Rellergert, Wade G. and Brown, Kenton R.},
  year = {2021},
  month = sep,
  journal = {Physical Review Letters},
  volume = {127},
  number = {13},
  eprint = {2105.05828},
  primaryclass = {physics, physics:quant-ph},
  pages = {130505},
  issn = {0031-9007, 1079-7114},
  doi = {10.1103/PhysRevLett.127.130505},
  urldate = {2024-08-12},
  archiveprefix = {arXiv},
  langid = {english}
}

@article{clarkPhaseNoiseMeasurements1999,
  title = {Phase Noise Measurements of Ultrastable 10 {{GHz}} Harmonically Modelocked Fibre Laser},
  author = {Clark, T.R. and Carruthers, T.F. and Matthews, P.J. and Duling, I.N.},
  year = {1999},
  month = apr,
  journal = {Electronics Letters},
  volume = {35},
  number = {9},
  pages = {720--721},
  publisher = {{The Institution of Engineering and Technology}},
  doi = {10.1049/el:19990516},
  urldate = {2025-06-04}
}

@article{duanScalingIonTrap2004,
  title = {Scaling {{Ion Trap Quantum Computation}} through {{Fast Quantum Gates}}},
  author = {Duan, L.-M.},
  year = {2004},
  month = sep,
  journal = {Physical Review Letters},
  volume = {93},
  number = {10},
  pages = {100502},
  issn = {0031-9007, 1079-7114},
  doi = {10.1103/PhysRevLett.93.100502},
  urldate = {2024-07-30},
  langid = {english}
}

@article{figgattParallelEntanglingOperations2019,
  title = {Parallel {{Entangling Operations}} on a {{Universal Ion Trap Quantum Computer}}},
  author = {Figgatt, C. and Ostrander, A. and Linke, N. M. and Landsman, K. A. and Zhu, D. and Maslov, D. and Monroe, C.},
  year = {2019},
  month = aug,
  journal = {Nature},
  volume = {572},
  number = {7769},
  eprint = {1810.11948},
  primaryclass = {quant-ph},
  pages = {368--372},
  issn = {0028-0836, 1476-4687},
  doi = {10.1038/s41586-019-1427-5},
  urldate = {2024-07-24},
  archiveprefix = {arXiv},
  langid = {english}
}

@article{gaeblerHighFidelityUniversalGate2016,
  title = {High-{{Fidelity Universal Gate Set}} for {{Be9}}+ {{Ion Qubits}}},
  author = {Gaebler, J. P. and Tan, T. R. and Lin, Y. and Wan, Y. and Bowler, R. and Keith, A. C. and Glancy, S. and Coakley, K. and Knill, E. and Leibfried, D. and Wineland, D. J.},
  year = {2016},
  month = aug,
  journal = {Physical Review Letters},
  volume = {117},
  number = {6},
  eprint = {1604.00032},
  pages = {060505},
  issn = {0031-9007, 1079-7114},
  doi = {10.1103/PhysRevLett.117.060505},
  urldate = {2024-03-31},
  archiveprefix = {arXiv}
}

@article{galeOptimizedFastGates2020,
  title = {Optimized Fast Gates for Quantum Computing with Trapped Ions},
  author = {Gale, Evan P. G. and Mehdi, Zain and Oberg, Lachlan M. and Ratcliffe, Alexander K. and Haine, Simon A. and Hope, Joseph J.},
  year = {2020},
  month = may,
  journal = {Physical Review A},
  volume = {101},
  number = {5},
  pages = {052328},
  publisher = {American Physical Society},
  doi = {10.1103/PhysRevA.101.052328},
  urldate = {2024-02-23}
}

@article{garcia-ripollCoherentControlTrapped2005,
  title = {Coherent Control of Trapped Ions Using Off-Resonant Lasers},
  author = {{Garcia-Ripoll}, J. J. and Zoller, P. and Cirac, J. I.},
  year = {2005},
  month = jun,
  journal = {Physical Review A},
  volume = {71},
  number = {6},
  eprint = {quant-ph/0411103},
  pages = {062309},
  issn = {1050-2947, 1094-1622},
  doi = {10.1103/PhysRevA.71.062309},
  urldate = {2024-02-23},
  archiveprefix = {arXiv}
}

@article{garcia-ripollSpeedOptimizedTwoQubit2003,
  title = {Speed {{Optimized Two-Qubit Gates}} with {{Laser Coherent Control Techniques}} for {{Ion Trap Quantum Computing}}},
  author = {{Garc{\'i}a-Ripoll}, Juan and Zoller, P and Cirac, J},
  year = {2003},
  month = nov,
  journal = {Physical review letters},
  volume = {91},
  pages = {157901},
  doi = {10.1103/PhysRevLett.91.157901}
}

@article{grzesiakEfficientArbitrarySimultaneously2020,
  title = {Efficient Arbitrary Simultaneously Entangling Gates on a Trapped-Ion Quantum Computer},
  author = {Grzesiak, Nikodem and Bl{\"u}mel, Reinhold and Wright, Kenneth and Beck, Kristin M. and Pisenti, Neal C. and Li, Ming and Chaplin, Vandiver and Amini, Jason M. and Debnath, Shantanu and Chen, Jwo-Sy and Nam, Yunseong},
  year = {2020},
  month = jun,
  journal = {Nature Communications},
  volume = {11},
  number = {1},
  pages = {2963},
  publisher = {Nature Publishing Group},
  issn = {2041-1723},
  doi = {10.1038/s41467-020-16790-9},
  urldate = {2024-08-04},
  copyright = {2020 The Author(s)},
  langid = {english}
}

@article{guoPicosecondIonqubitManipulation2022,
  title = {Picosecond Ion-Qubit Manipulation and Spin-Phonon Entanglement with Resonant Laser Pulses},
  author = {Guo, W.-X. and Wu, Y.-K. and Huang, Y.-Y. and Feng, L. and Huang, C.-X. and Yang, H.-X. and Ma, J.-Y. and Yao, L. and Zhou, Z.-C. and Duan, L.-M.},
  year = {2022},
  month = aug,
  journal = {Physical Review A},
  volume = {106},
  number = {2},
  pages = {022608},
  issn = {2469-9926, 2469-9934},
  doi = {10.1103/PhysRevA.106.022608},
  urldate = {2024-09-03},
  langid = {english}
}

@article{hartyHighfidelityPreparationGates2014,
  title = {High-Fidelity Preparation, Gates, Memory and Readout of a Trapped-Ion Quantum Bit},
  author = {Harty, T. P. and Allcock, D. T. C. and Ballance, C. J. and Guidoni, L. and Janacek, H. A. and Linke, N. M. and Stacey, D. N. and Lucas, D. M.},
  year = {2014},
  month = nov,
  journal = {Physical Review Letters},
  volume = {113},
  number = {22},
  eprint = {1403.1524},
  primaryclass = {quant-ph},
  pages = {220501},
  issn = {0031-9007, 1079-7114},
  doi = {10.1103/PhysRevLett.113.220501},
  urldate = {2024-04-01},
  archiveprefix = {arXiv}
}

@article{hayesEntanglementAtomicQubits2010,
  title = {Entanglement of {{Atomic Qubits Using}} an {{Optical Frequency Comb}}},
  author = {Hayes, D. and Matsukevich, D. N. and Maunz, P. and Hucul, D. and Quraishi, Q. and Olmschenk, S. and Campbell, W. and Mizrahi, J. and Senko, C. and Monroe, C.},
  year = {2010},
  month = apr,
  journal = {Physical Review Letters},
  volume = {104},
  number = {14},
  pages = {140501},
  issn = {0031-9007, 1079-7114},
  doi = {10.1103/PhysRevLett.104.140501},
  urldate = {2024-09-03},
  langid = {english}
}

@article{heinrichUltrafastCoherentExcitation2019,
  title = {Ultrafast Coherent Excitation of a {{40Ca}}+ Ion},
  author = {Heinrich, D. and Guggemos, M. and {Guevara-Bertsch}, M. and Hussain, M. I. and Roos, C. F. and Blatt, R.},
  year = {2019},
  month = jul,
  journal = {New Journal of Physics},
  volume = {21},
  number = {7},
  pages = {073017},
  publisher = {IOP Publishing},
  issn = {1367-2630},
  doi = {10.1088/1367-2630/ab2a7e},
  urldate = {2024-08-26},
  langid = {english}
}

@article{homeNormalModesTrapped2011,
  title = {Normal Modes of Trapped Ions in the Presence of Anharmonic Trap Potentials},
  author = {Home, J P and Hanneke, D and Jost, J D and Leibfried, D and Wineland, D J},
  year = {2011},
  month = jul,
  journal = {New Journal of Physics},
  volume = {13},
  number = {7},
  pages = {073026},
  issn = {1367-2630},
  doi = {10.1088/1367-2630/13/7/073026},
  urldate = {2024-08-05},
  langid = {english}
}

@article{hussainUltrafastHighRepetition2016,
  title = {Ultrafast, High Repetition Rate, Ultraviolet, Fiber-Laser-Based Source: Application towards {{Yb}}{\textsuperscript{+}} Fast Quantum-Logic},
  shorttitle = {Ultrafast, High Repetition Rate, Ultraviolet, Fiber-Laser-Based Source},
  author = {Hussain, Mahmood Irtiza and Petrasiunas, Matthew Joseph and Bentley, Christopher D. B. and Taylor, Richard L. and Carvalho, Andr{\'e} R. R. and Hope, Joseph J. and Streed, Erik W. and Lobino, Mirko and Kielpinski, David},
  year = {2016},
  month = jul,
  journal = {Optics Express},
  volume = {24},
  number = {15},
  pages = {16638--16648},
  publisher = {Optica Publishing Group},
  issn = {1094-4087},
  doi = {10.1364/OE.24.016638},
  urldate = {2024-08-25},
  copyright = {{\copyright} 2016 Optical Society of America},
  langid = {english}
}

@article{jamesQuantumDynamicsCold1998,
  title = {Quantum Dynamics of Cold Trapped Ions, with Application to Quantum Computation},
  author = {James, Daniel F. V.},
  year = {1998},
  month = feb,
  journal = {Applied Physics B: Lasers and Optics},
  volume = {66},
  number = {2},
  eprint = {quant-ph/9702053},
  pages = {181--190},
  issn = {0946-2171, 1432-0649},
  doi = {10.1007/s003400050373},
  urldate = {2024-08-05},
  archiveprefix = {arXiv},
  langid = {english}
}

@article{johnsonUltrafastCreationLarge2017,
  title = {Ultrafast Creation of Large {{Schr{\"o}dinger}} Cat States of an Atom},
  author = {Johnson, K. G. and {Wong-Campos}, J. D. and Neyenhuis, B. and Mizrahi, J. and Monroe, C.},
  year = {2017},
  month = sep,
  journal = {Nature Communications},
  volume = {8},
  number = {1},
  pages = {697},
  publisher = {Nature Publishing Group},
  issn = {2041-1723},
  doi = {10.1038/s41467-017-00682-6},
  urldate = {2025-06-23},
  copyright = {2017 The Author(s)},
  langid = {english}
}

@article{kaushalShuttlingbasedTrappedionQuantum2020,
  title = {Shuttling-Based Trapped-Ion Quantum Information Processing},
  author = {Kaushal, V. and Lekitsch, B. and Stahl, A. and Hilder, J. and Pijn, D. and Schmiegelow, C. and Bermudez, A. and M{\"u}ller, M. and {Schmidt-Kaler}, F. and Poschinger, U.},
  year = {2020},
  month = mar,
  journal = {AVS Quantum Science},
  volume = {2},
  number = {1},
  pages = {014101},
  issn = {2639-0213},
  doi = {10.1116/1.5126186},
  urldate = {2025-06-30}
}

@misc{knollmannIntegratedPhotonicStructures2024,
  title = {Integrated Photonic Structures for Photon-Mediated Entanglement of Trapped Ions},
  author = {Knollmann, F. W. and Clements, E. and Callahan, P. T. and Gehl, M. and Hunker, J. D. and Mahony, T. and McConnell, R. and Swint, R. and {Sorace-Agaskar}, C. and Chuang, I. L. and Chiaverini, J. and Stick, D.},
  year = {2024},
  month = jan,
  number = {arXiv:2401.06850},
  eprint = {2401.06850},
  primaryclass = {physics, physics:quant-ph},
  publisher = {arXiv},
  urldate = {2024-09-25},
  archiveprefix = {arXiv},
  langid = {english}
}

@article{landsmanTwoqubitEntanglingGates2019,
  title = {Two-Qubit Entangling Gates within Arbitrarily Long Chains of Trapped Ions},
  author = {Landsman, K. A. and Wu, Y. and Leung, P. H. and Zhu, D. and Linke, N. M. and Brown, K. R. and Duan, L. and Monroe, C.},
  year = {2019},
  month = aug,
  journal = {Physical Review A},
  volume = {100},
  number = {2},
  pages = {022332},
  issn = {2469-9926, 2469-9934},
  doi = {10.1103/PhysRevA.100.022332},
  urldate = {2024-08-05},
  langid = {english}
}

@article{leungEntanglingArbitraryPair2018,
  title = {Entangling an Arbitrary Pair of Qubits in a Long Ion Crystal},
  author = {Leung, Pak Hong and Brown, Kenneth R.},
  year = {2018},
  month = sep,
  journal = {Physical Review A},
  volume = {98},
  number = {3},
  eprint = {1808.02555},
  primaryclass = {physics, physics:quant-ph},
  pages = {032318},
  issn = {2469-9926, 2469-9934},
  doi = {10.1103/PhysRevA.98.032318},
  urldate = {2024-08-12},
  archiveprefix = {arXiv},
  langid = {english}
}

@article{liHomogeneousLinearIon2022,
  title = {Homogeneous {{Linear Ion Crystal}} in a {{Hybrid Potential}}},
  author = {Li, Ming-shen and Liu, Yang and Rao, Xin-Xin and Lu, Peng-Fei and Wang, Zhao and Zhu, Feng and Luo, Le},
  year = {2022},
  month = feb,
  journal = {Quantum Information Processing},
  volume = {21},
  number = {2},
  eprint = {2103.02158},
  primaryclass = {physics},
  pages = {65},
  issn = {1570-0755, 1573-1332},
  doi = {10.1007/s11128-022-03412-0},
  urldate = {2024-08-06},
  archiveprefix = {arXiv}
}

@article{linLargescaleQuantumComputation2009,
  title = {Large-Scale Quantum Computation in an Anharmonic Linear Ion Trap},
  author = {Lin, G.-D. and Zhu, S.-L. and Islam, R. and Kim, K. and Chang, M.-S. and Korenblit, S. and Monroe, C. and Duan, L.-M.},
  year = {2009},
  month = jul,
  journal = {Europhysics Letters},
  volume = {86},
  number = {6},
  pages = {60004},
  issn = {0295-5075},
  doi = {10.1209/0295-5075/86/60004},
  urldate = {2024-07-11},
  langid = {english}
}

@misc{loschnauerScalableHighfidelityAllelectronic2024,
  title = {Scalable, High-Fidelity All-Electronic Control of Trapped-Ion Qubits},
  author = {L{\"o}schnauer, C. M. and Toba, J. Mosca and Hughes, A. C. and King, S. A. and Weber, M. A. and Srinivas, R. and Matt, R. and Nourshargh, R. and Allcock, D. T. C. and Ballance, C. J. and Matthiesen, C. and Malinowski, M. and Harty, T. P.},
  year = {2024},
  month = jul,
  number = {arXiv:2407.07694},
  eprint = {2407.07694},
  primaryclass = {quant-ph},
  publisher = {arXiv},
  doi = {10.48550/arXiv.2407.07694},
  urldate = {2025-06-28},
  archiveprefix = {arXiv}
}

@misc{mainDistributedQuantumComputing2024,
  title = {Distributed {{Quantum Computing}} across an {{Optical Network Link}}},
  author = {Main, D. and Drmota, P. and Nadlinger, D. P. and Ainley, E. M. and Agrawal, A. and Nichol, B. C. and Srinivas, R. and Araneda, G. and Lucas, D. M.},
  year = {2024},
  month = jun,
  number = {arXiv:2407.00835},
  eprint = {2407.00835},
  primaryclass = {quant-ph},
  publisher = {arXiv},
  urldate = {2024-09-25},
  archiveprefix = {arXiv},
  langid = {english}
}

@misc{maiScalableEntanglingGates2025,
  title = {Scalable Entangling Gates on Ion Qubits via Structured Light Addressing},
  author = {Mai, Xueying and Zhang, Liyun and Yu, Qinyang and Zhang, Junhua and Lu, Yao},
  year = {2025},
  month = jun,
  number = {arXiv:2506.19535},
  eprint = {2506.19535},
  primaryclass = {quant-ph},
  publisher = {arXiv},
  doi = {10.48550/arXiv.2506.19535},
  urldate = {2025-06-28},
  archiveprefix = {arXiv}
}

@article{manovitzTrappedIonQuantumComputer2022,
  title = {Trapped-{{Ion Quantum Computer}} with {{Robust Entangling Gates}} and {{Quantum Coherent Feedback}}},
  author = {Manovitz, Tom and Shapira, Yotam and Gazit, Lior and Akerman, Nitzan and Ozeri, Roee},
  year = {2022},
  month = mar,
  journal = {PRX Quantum},
  volume = {3},
  number = {1},
  pages = {010347},
  issn = {2691-3399},
  doi = {10.1103/PRXQuantum.3.010347},
  urldate = {2024-08-07},
  langid = {english}
}

@article{mehdiFastEntanglingGates2021,
  title = {Fast Entangling Gates in Long Ion Chains},
  author = {Mehdi, Zain and Ratcliffe, Alexander K. and Hope, Joseph J.},
  year = {2021},
  month = jan,
  journal = {Physical Review Research},
  volume = {3},
  number = {1},
  pages = {013026},
  publisher = {American Physical Society},
  doi = {10.1103/PhysRevResearch.3.013026},
  urldate = {2024-02-23}
}

@misc{mehdiFastMixedspeciesQuantum2025,
  title = {Fast Mixed-Species Quantum Logic Gates for Trapped-Ion Quantum Networks},
  author = {Mehdi, Zain and Vaidya, Varun D. and {Savill-Brown}, Isabelle and Grosser, Phoebe and Ratcliffe, Alexander K. and Liu, Haonan and Haine, Simon A. and Hope, Joseph J. and Viteri, C. Ricardo},
  year = {2025},
  month = mar,
  number = {arXiv:2412.07185},
  eprint = {2412.07185},
  primaryclass = {quant-ph},
  publisher = {arXiv},
  doi = {10.48550/arXiv.2412.07185},
  urldate = {2025-05-26},
  archiveprefix = {arXiv}
}

@article{mehdiScalableQuantumComputation2020,
  title = {Scalable Quantum Computation with Fast Gates in Two-Dimensional Microtrap Arrays of Trapped Ions},
  author = {Mehdi, Zain and Ratcliffe, Alexander K. and Hope, Joseph J.},
  year = {2020},
  month = jul,
  journal = {Physical Review A},
  volume = {102},
  number = {1},
  pages = {012618},
  publisher = {American Physical Society},
  doi = {10.1103/PhysRevA.102.012618},
  urldate = {2024-02-23}
}

@article{mizrahiQuantumControlQubits2014,
  title = {Quantum Control of Qubits and Atomic Motion Using Ultrafast Laser Pulses},
  author = {Mizrahi, J. and Neyenhuis, B. and Johnson, K. G. and Campbell, W. C. and Senko, C. and Hayes, D. and Monroe, C.},
  year = {2014},
  month = jan,
  journal = {Applied Physics B},
  volume = {114},
  number = {1},
  pages = {45--61},
  issn = {1432-0649},
  doi = {10.1007/s00340-013-5717-6},
  urldate = {2024-04-21},
  langid = {english}
}

@article{mizrahiUltrafastSpinMotionEntanglement2013,
  title = {Ultrafast {{Spin-Motion Entanglement}} and {{Interferometry}} with a {{Single Atom}}},
  author = {Mizrahi, J. and Senko, C. and Neyenhuis, B. and Johnson, K. G. and Campbell, W. C. and Conover, C. W. S. and Monroe, C.},
  year = {2013},
  month = may,
  journal = {Physical Review Letters},
  volume = {110},
  number = {20},
  pages = {203001},
  issn = {0031-9007, 1079-7114},
  doi = {10.1103/PhysRevLett.110.203001},
  urldate = {2024-10-14},
  copyright = {http://link.aps.org/licenses/aps-default-license},
  langid = {english}
}

@article{monroeLargescaleModularQuantumcomputer2014,
  title = {Large-Scale Modular Quantum-Computer Architecture with Atomic Memory and Photonic Interconnects},
  author = {Monroe, C. and Raussendorf, R. and Ruthven, A. and Brown, K. R. and Maunz, P. and Duan, L.-M. and Kim, J.},
  year = {2014},
  month = feb,
  journal = {Physical Review A},
  volume = {89},
  number = {2},
  pages = {022317},
  issn = {1050-2947, 1094-1622},
  doi = {10.1103/PhysRevA.89.022317},
  urldate = {2024-09-06},
  copyright = {http://link.aps.org/licenses/aps-default-license},
  langid = {english}
}

@article{mosesRaceTrackTrappedIonQuantum2023,
  title = {A {{Race-Track Trapped-Ion Quantum Processor}}},
  author = {Moses, S. A. and Baldwin, C. H. and Allman, M. S. and Ancona, R. and Ascarrunz, L. and Barnes, C. and Bartolotta, J. and Bjork, B. and Blanchard, P. and Bohn, M. and Bohnet, J. G. and Brown, N. C. and Burdick, N. Q. and Burton, W. C. and Campbell, S. L. and Campora, J. P. and Carron, C. and Chambers, J. and Chan, J. W. and Chen, Y. H. and Chernoguzov, A. and Chertkov, E. and Colina, J. and Curtis, J. P. and Daniel, R. and DeCross, M. and Deen, D. and Delaney, C. and Dreiling, J. M. and Ertsgaard, C. T. and Esposito, J. and Estey, B. and Fabrikant, M. and Figgatt, C. and Foltz, C. and {Foss-Feig}, M. and Francois, D. and Gaebler, J. P. and Gatterman, T. M. and Gilbreth, C. N. and Giles, J. and Glynn, E. and Hall, A. and Hankin, A. M. and Hansen, A. and Hayes, D. and Higashi, B. and Hoffman, I. M. and Horning, B. and Hout, J. J. and Jacobs, R. and Johansen, J. and Jones, L. and Karcz, J. and Klein, T. and Lauria, P. and Lee, P. and Liefer, D. and Lu, S. T. and Lucchetti, D. and Lytle, C. and Malm, A. and Matheny, M. and Mathewson, B. and Mayer, K. and Miller, D. B. and Mills, M. and Neyenhuis, B. and Nugent, L. and Olson, S. and Parks, J. and Price, G. N. and Price, Z. and Pugh, M. and Ransford, A. and Reed, A. P. and Roman, C. and Rowe, M. and {Ryan-Anderson}, C. and Sanders, S. and Sedlacek, J. and Shevchuk, P. and Siegfried, P. and Skripka, T. and Spaun, B. and Sprenkle, R. T. and Stutz, R. P. and Swallows, M. and Tobey, R. I. and Tran, A. and Tran, T. and Vogt, E. and Volin, C. and Walker, J. and Zolot, A. M. and Pino, J. M.},
  year = {2023},
  month = dec,
  journal = {Physical Review X},
  volume = {13},
  number = {4},
  pages = {041052},
  issn = {2160-3308},
  doi = {10.1103/PhysRevX.13.041052},
  urldate = {2024-09-03},
  langid = {english}
}

@article{paganoCryogenicTrappedIonSystem2018,
  title = {Cryogenic {{Trapped-Ion System}} for {{Large Scale Quantum Simulation}}},
  author = {Pagano, G. and Hess, P. W. and Kaplan, H. B. and Tan, W. L. and Richerme, P. and Becker, P. and Kyprianidis, A. and Zhang, J. and Birckelbaw, E. and Hernandez, M. R. and Wu, Y. and Monroe, C.},
  year = {2018},
  month = oct,
  journal = {Quantum Science and Technology},
  volume = {4},
  number = {1},
  eprint = {1802.03118},
  primaryclass = {physics, physics:quant-ph},
  pages = {014004},
  issn = {2058-9565},
  doi = {10.1088/2058-9565/aae0fe},
  urldate = {2024-09-03},
  archiveprefix = {arXiv},
  langid = {english}
}

@article{pinoDemonstrationTrappedionQuantum2021,
  title = {Demonstration of the Trapped-Ion Quantum {{CCD}} Computer Architecture},
  author = {Pino, J. M. and Dreiling, J. M. and Figgatt, C. and Gaebler, J. P. and Moses, S. A. and Allman, M. S. and Baldwin, C. H. and {Foss-Feig}, M. and Hayes, D. and Mayer, K. and {Ryan-Anderson}, C. and Neyenhuis, B.},
  year = {2021},
  month = apr,
  journal = {Nature},
  volume = {592},
  number = {7853},
  pages = {209--213},
  publisher = {Nature Publishing Group},
  issn = {1476-4687},
  doi = {10.1038/s41586-021-03318-4},
  urldate = {2024-09-23},
  copyright = {2021 The Author(s), under exclusive licence to Springer Nature Limited part of Springer Nature},
  langid = {english}
}

@article{putnamImpulsiveSpinmotionEntanglement2024,
  title = {Impulsive Spin-Motion Entanglement for Fast Quantum Computation and Sensing},
  author = {Putnam, Randall and West, Adam D. and Campbell, Wesley C. and Hamilton, Paul},
  year = {2024},
  month = mar,
  journal = {Physical Review A},
  volume = {109},
  number = {3},
  pages = {032614},
  issn = {2469-9926, 2469-9934},
  doi = {10.1103/PhysRevA.109.032614},
  urldate = {2025-06-28},
  langid = {english}
}

@article{ratcliffeMicromotionenhancedFastEntangling2020,
  title = {Micromotion-Enhanced Fast Entangling Gates for Trapped-Ion Quantum Computing},
  author = {Ratcliffe, Alexander K. and Oberg, Lachlan M. and Hope, Joseph J.},
  year = {2020},
  month = may,
  journal = {Physical Review A},
  volume = {101},
  number = {5},
  pages = {052332},
  publisher = {American Physical Society},
  doi = {10.1103/PhysRevA.101.052332},
  urldate = {2024-02-23}
}

@article{ratcliffeScalingTrappedIon2018,
  title = {Scaling {{Trapped Ion Quantum Computers Using Fast Gates}} and {{Microtraps}}},
  author = {Ratcliffe, Alexander K. and Taylor, Richard L. and Hope, Joseph J. and Carvalho, Andr{\'e} R. R.},
  year = {2018},
  month = may,
  journal = {Physical Review Letters},
  volume = {120},
  number = {22},
  pages = {220501},
  publisher = {American Physical Society},
  doi = {10.1103/PhysRevLett.120.220501},
  urldate = {2024-10-06}
}

@article{sanerBreakingEntanglingGate2023,
  title = {Breaking the {{Entangling Gate Speed Limit}} for {{Trapped-Ion Qubits Using}} a {{Phase-Stable Standing Wave}}},
  author = {Saner, S. and B{\u a}z{\u a}van, O. and Minder, M. and Drmota, P. and Webb, D. J. and Araneda, G. and Srinivas, R. and Lucas, D. M. and Ballance, C. J.},
  year = {2023},
  month = dec,
  journal = {Physical Review Letters},
  volume = {131},
  number = {22},
  pages = {220601},
  publisher = {American Physical Society},
  doi = {10.1103/PhysRevLett.131.220601},
  urldate = {2024-09-17}
}

@article{savill-brownHighspeedHighconnectivityTwoqubit2025,
  title = {High-{{Speed}} and {{High-Connectivity Two-Qubit Gates}} in {{Long Chains}} of {{Trapped Ions}}},
  author = {{Savill-Brown}, Isabelle and Hope, Joseph J. and Ratcliffe, Alexander K. and Vaidya, Varun D. and Liu, Haonan and Haine, Simon A. and Viteri, C. Ricardo and Mehdi, Zain},
  year = 2026,
  month = may,
  journal = {Physical Review Letters},
  volume = {136},
  number = {19},
  pages = {190802},
  publisher = {American Physical Society},
  doi = {10.1103/45zd-f4my},
  urldate = {2026-05-15}
}

@article{schaferFastQuantumLogic2018,
  title = {Fast Quantum Logic Gates with Trapped-Ion Qubits},
  author = {Sch{\"a}fer, V. M. and Ballance, C. J. and Thirumalai, K. and Stephenson, L. J. and Ballance, T. G. and Steane, A. M. and Lucas, D. M.},
  year = {2018},
  month = mar,
  journal = {Nature},
  volume = {555},
  number = {7694},
  pages = {75--78},
  publisher = {Nature Publishing Group},
  issn = {1476-4687},
  doi = {10.1038/nature25737},
  urldate = {2024-02-27},
  copyright = {2018 Macmillan Publishers Limited, part of Springer Nature. All rights reserved.},
  langid = {english}
}

@article{sorensenEntanglementQuantumComputation2000,
  title = {Entanglement and Quantum Computation with Ions in Thermal Motion},
  author = {Sorensen, Anders and Molmer, Klaus},
  year = {2000},
  month = jul,
  journal = {Physical Review A},
  volume = {62},
  number = {2},
  eprint = {quant-ph/0002024},
  pages = {022311},
  issn = {1050-2947, 1094-1622},
  doi = {10.1103/PhysRevA.62.022311},
  urldate = {2024-08-12},
  archiveprefix = {arXiv},
  langid = {english}
}

@article{steanePulsedForceSequences2014,
  title = {Pulsed Force Sequences for Fast Phase-Insensitive Quantum Gates in Trapped Ions},
  author = {Steane, A M and Imreh, G and Home, J P and Leibfried, D},
  year = {2014},
  month = may,
  journal = {New Journal of Physics},
  volume = {16},
  number = {5},
  pages = {053049},
  issn = {1367-2630},
  doi = {10.1088/1367-2630/16/5/053049},
  urldate = {2024-03-02},
  langid = {english}
}

@article{stephensonHighrateHighfidelityEntanglement2020,
  title = {High-Rate, High-Fidelity Entanglement of Qubits across an Elementary Quantum Network},
  author = {Stephenson, L. J. and Nadlinger, D. P. and Nichol, B. C. and An, S. and Drmota, P. and Ballance, T. G. and Thirumalai, K. and Goodwin, J. F. and Lucas, D. M. and Ballance, C. J.},
  year = {2020},
  month = mar,
  journal = {Physical Review Letters},
  volume = {124},
  number = {11},
  eprint = {1911.10841},
  primaryclass = {physics, physics:quant-ph},
  pages = {110501},
  issn = {0031-9007, 1079-7114},
  doi = {10.1103/PhysRevLett.124.110501},
  urldate = {2024-03-11},
  archiveprefix = {arXiv}
}

@article{talukdarImplicationsSurfaceNoise2016,
  title = {Implications of Surface Noise for the Motional Coherence of Trapped Ions},
  author = {Talukdar, I. and Gorman, D. J. and Daniilidis, N. and Schindler, P. and Ebadi, S. and Kaufmann, H. and Zhang, T. and H{\"a}ffner, H.},
  year = {2016},
  month = apr,
  journal = {Physical Review A},
  volume = {93},
  number = {4},
  pages = {043415},
  publisher = {American Physical Society},
  doi = {10.1103/PhysRevA.93.043415},
  urldate = {2025-06-04}
}

@article{taylorStudyFastGates2017,
  title = {A {{Study}} on {{Fast Gates}} for {{Large-Scale Quantum Simulation}} with {{Trapped Ions}}},
  author = {Taylor, Richard L. and Bentley, Christopher D. B. and Pedernales, Julen S. and Lamata, Lucas and Solano, Enrique and Carvalho, Andr{\'e} R. R. and Hope, Joseph J.},
  year = {2017},
  month = apr,
  journal = {Scientific Reports},
  volume = {7},
  number = {1},
  pages = {46197},
  publisher = {Nature Publishing Group},
  issn = {2045-2322},
  doi = {10.1038/srep46197},
  urldate = {2024-02-23},
  langid = {english}
}

@article{wangSingleIonqubitExceeding2021,
  title = {Single Ion-Qubit Exceeding One Hour Coherence Time},
  author = {Wang, Pengfei and Luan, Chun-Yang and Qiao, Mu and Um, Mark and Zhang, Junhua and Wang, Ye and Yuan, Xiao and Gu, Mile and Zhang, Jingning and Kim, Kihwan},
  year = {2021},
  month = jan,
  journal = {Nature Communications},
  volume = {12},
  number = {1},
  eprint = {2008.00251},
  primaryclass = {quant-ph},
  pages = {233},
  issn = {2041-1723},
  doi = {10.1038/s41467-020-20330-w},
  urldate = {2024-04-11},
  archiveprefix = {arXiv}
}

@article{wimperisBroadbandNarrowbandPassband1994,
  title = {Broadband, {{Narrowband}}, and {{Passband Composite Pulses}} for {{Use}} in {{Advanced NMR Experiments}}},
  author = {Wimperis, S.},
  year = {1994},
  month = aug,
  journal = {Journal of Magnetic Resonance, Series A},
  volume = {109},
  number = {2},
  pages = {221--231},
  issn = {1064-1858},
  doi = {10.1006/jmra.1994.1159},
  urldate = {2024-02-23}
}

@article{wong-camposDemonstrationTwoAtomEntanglement2017,
  title = {Demonstration of {{Two-Atom Entanglement}} with {{Ultrafast Optical Pulses}}},
  author = {{Wong-Campos}, J. D. and Moses, S. A. and Johnson, K. G. and Monroe, C.},
  year = {2017},
  month = dec,
  journal = {Physical Review Letters},
  volume = {119},
  number = {23},
  pages = {230501},
  publisher = {American Physical Society},
  doi = {10.1103/PhysRevLett.119.230501},
  urldate = {2025-06-24}
}

@article{mizrahi_ultrafast_2013,
    title = {Ultrafast {Spin}-{Motion} {Entanglement} and {Interferometry} with a {Single} {Atom}},
    volume = {110},
    url = {https://link.aps.org/doi/10.1103/PhysRevLett.110.203001},
    doi = {10.1103/PhysRevLett.110.203001},
    number = {20},
    urldate = {2024-09-09},
    journal = {Physical Review Letters},
    author = {Mizrahi, J. and Senko, C. and Neyenhuis, B. and Johnson, K. G. and Campbell, W. C. and Conover, C. W. S. and Monroe, C.},
    month = may,
    year = {2013},
    note = {Publisher: American Physical Society},
    pages = {203001},
}

\end{document}